\documentclass[twocolumn,showpacs,preprintnumbers,amsmath,amssymb,prb]{revtex4-1}
\usepackage{graphicx}
\usepackage{dcolumn}
\usepackage{bm}

\begin{document}

\newcommand{\be}{\begin{eqnarray}}
\newcommand{\ee}{\end{eqnarray}}
\newcommand{\nn}{\nonumber\\}
\newcommand{\nin}{\noindent}
\newcommand{\la}{\langle}
\newcommand{\ra}{\rangle}

\renewcommand{\theequation}{\arabic{section}.\arabic{equation}}
\newcommand{\Ub}{\bar{U}}
\newcommand{\Db}{\bar{D}}
\newcommand{\zb}{\bar{z}}
\newcommand{\pb}{\bar{p}}
\newcommand{\omegab}{\bar{\omega}}
\newcommand{\wib}{\bar{w}_{i}}
\newcommand{\wi}{{w}_{i}}
\newcommand{\wjb}{\bar{w}_{j}}
\newcommand{\wj}{{w}_{j}}
\newcommand{\wb}{\bar{w}_{1}}
\newcommand{\w}{{w}_{1}}
\newcommand{\Jm}{\mathcal{J}}
\newcommand{\Dm}{\mathcal{D}}
\newcommand{\dellr}{\overleftrightarrow{\partial}}
\newcommand{\Tr}{\mathrm{Tr}}

\title{Lattice Ginzburg-Landau Model of a Ferromagnetic $p$-wave Pairing
Phase in Superconducting Materials  and an Inhomogeneous Coexisting State
}

\author{Akihiro Shimizu, Hidetoshi Ozawa, Ikuo Ichinose}
\affiliation{Department of Applied Physics, Nagoya Institute of Technology,
Nagoya, 466-8555 Japan}
\author{Tetsuo Matsui}
\affiliation{Department of Physics, Kinki University,
Higashi-Osaka, 577-8502 Japan}

\date{\today}

\begin{abstract}
We study the interplay of the
ferromagnetic (FM) state and the $p$-wave superconducting (SC) state
observed in several materials such as UCoGe and URhGe in a
totally nonperturbative manner.
To this end, we  introduce a lattice Ginzburg-Landau model 
that is a genuine generalization of the phenomenological
Ginzburg-Landau theory proposed previously in the continuum and also a 
counterpart of the lattice gauge-Higgs model for the $s$-wave SC transition,
and study it numerically by Monte-Carlo simulations.
The obtained phase diagram 
has qualitatively the same structure as that of UCoGe
in the region where the two transition temperatrures satisfy 
$T_{\rm FM}>T_{\rm SC}$.
For $T_{\rm FM}/T_{\rm SC} < 0.7$, we find that
the coexisting region of  FM and SC orders appears only near the 
surface of the lattice, which describes an
inhomogeneous FMSC coexisting state.
\end{abstract}

\maketitle

\section{Introduction}
In the last decade, superconducting (SC) materials 
coexisting with ferromagnetic (FM) long-range orders have been found
and of intensive interest.
In UGe$_2$\cite{UG} and URhGe\cite{UR},
a SC state appears only within the FM state in the pressure-temperature
($P$-$T$) phase diagram, whereas
in UCoGe\cite{UC}, the SC state exists both in the FM and 
paramagnetic states\cite{UC2}.
Phenomenological models of FMSC materials were 
proposed\cite{MO,mineev} soon after their discovery.
The most important observation in those studies is that the FMSC state is
a spin-triplet $p$-wave state of electron pairs\cite{Harada}. 

In the present paper, we propose a lattice model for describing FMSC
materials and investigate it by numerical methods\cite{LT26}.
The model contains a vector potential (i.e., gauge field) to describe
a FM order parameter(magnetization) 
and also a SC order parameter, i.e., Cooper-pair field for the $p$-wave 
SC state.
These two physical variables couple with each other as the Cooper pair
bears the electric charge $2e$.
Finite magnetization inside the material tends to induce vortices of the
Cooper-pair field and destabilize SC state.
In this sense, the FMSC state is a result of frustration. 

Introduction of the spatial lattice in  the present model
has several advantages over the Ginzburg-Landau (GL) theory 
in the continuum space\cite{MO,mineev,GL};
(i) it reflects the lattice structure of the real 
materials, (ii) it works as a reguralization of vortex configurations because 
the energy of these topological excitations becomes finite,  and (iii)
it allows us to make a nonperturbative study by means of Monte-Carlo 
simulations in which contributions from all the field configurations 
including topologically nontrivial excitations
are taken into account, and so
the obtained results are quite reliable.
In this sense, the present study is complementary to the previous
analytical studies employing perturbative and mean-field like
approximations\cite{MO,mineev,GL}.

The present paper is organized as follows.
In Sec.II, we introduce the lattice GL model that is derived 
the previously propose GL theory in the continuum.
Detailed discussion on the physical properties of the model is also given 
there. 
In Sect.II.A, we present a brief review on the lattice gauge model for 
the SC state.
This review may be useful to make the present paper readable
for condensed matter physicists who are not familiar with the models 
of SC state on the lattice.
In Sec.III, we present results of the numerical study.
The phase structure of the model is clarified by calculating specific heat,
FM correlation function, shielding mass of magnetic field, etc.
We also study behavior of vortices in a constant magnetic field in the 
present model in order to obtain an intuitive picture of the Meissner state.
Section IV is devoted for conclusion.

\section{Lattice Model for FMSC State}
\setcounter{equation}{0}
\subsection{Lattice gauge-Higgs model for  SC transition}
In this subsection, we review a typical lattice model 
to describe the conventional SC transition,
which is called the U(1) gauge-Higgs model (or the Abelian Higgs model).
Reader who is familiar with this subject can skip this subsection and
go to Sec.II.B.

Let us start with the GL theory of $s$-wave SC state 
in the three-dimensional (3D) continuum space. 
Its  free-energy density is given by 
\be
f_{s{\rm GL}}&=&|D_\mu\psi|^2+
\alpha(T-T_{\rm{c}}^0)|\psi|^2
+\lambda|\psi|^4+\frac{1}{8e^2}({\rm rot}\vec{A}^{\rm em})^2,\nn
D_\mu&=&\partial_\mu-iA_\mu^{\rm em},
\label{sGL}
\ee 
where $\psi$ is the  complex scalar field for $s$-wave Cooper pairs, 
$\vec{A}^{\rm em}$ is the vector potential ($\times 2e$) for fluctuating 
electro-magnetic field,
$D_\mu$ is the covariant derivative in the $\mu$-th direction ($\mu=1,2,3$),
$T_{\rm{c}}^0$ is the {\em bare} critical temperature($T$) of SC transition,
$e$ is the elementary charge, 
$\alpha$ and $\lambda$ are positive $T$-independent parameters.

The lattice-field model corresponding to the GL theory (\ref{sGL}) 
is defined by giving its free energy (or the action including the
inverse temperature) $F$ as follows;
\be
&& F=-\frac{K}{2}\sum_{x,\mu}\left(\psi^\ast_{x+\mu}U_{x\mu}\psi_x+{\rm c.c.}\right)
+F_\psi+F_A,\label{LGT1} \\
&& F_\psi=\sum_x\left(\sigma|\psi_x|^2
+\lambda|\psi_x|^4\right),\label{Fpsi}  \\
&& U_{x\mu}= \exp(iA_{x\mu}^{\rm em}), 
\nonumber
\ee
where $x$ is the site of the 3D lattice, $\psi_x$ is a complex 
SC order-parameter field defined on the site $x$, and 
we use $\mu$ also as the unit vector in the $\mu$-th direction.
$\psi_x$ is sometimes called  Higgs field because it
is a complex scalar field.
$A_{x\mu}^{\rm em}$ is a real electro-magnetic 
field put on the link $(x,x+\mu)$.
$F_A$ is the free energy  of the electro-magnetic field and 
has the following two versions. 
One is the compact version,  
\be
F_A&=&-\frac{1}{2e^2}\sum_{x,\mu < \nu}\left(
U^\ast_{x\nu}U^\ast_{x+\nu,\mu}U_{x+\mu,\nu}U_{x\mu} +{\rm c.c.}\right),\nn
A_{x\mu}^{\rm em}&\in& (-\pi,\pi),
\label{FA1}
\ee 
The other is the noncompact version,
\be
F_A&=&\frac{1}{2e^2}\sum_{x,\mu < \nu}\left(
\nabla_\mu A_{x\nu}^{\rm em}-\nabla_\nu A_{x\mu}^{\rm em}\right)^2,\nn
A_{x\mu}^{\rm em}&\in& (-\infty,\infty),
\label{FA2}
\ee 
where $\nabla_\mu$ is the lattice difference operator such that 
\be
\nabla_\mu f_x \equiv f_{x+\mu}-f_x.
\ee
These two $F_A$  are distinguished by having
the periodicity under $A_{x\mu}\to A_{x\mu}+2\pi$ or not.
We note that $F$ of (\ref{LGT1}) is invariant under the local U(1)
gauge transformation,
\be
\psi_x &\to& \psi'_x =\exp(i\lambda_x)\psi_x,\nn
A_{x\mu}^{\rm em} &\to& A_{x\mu}^{'{\rm em}} =A_{x\mu}^{\rm em}
+\lambda_{x+\mu}-\lambda_x,\nn
U_{x\mu}&\to&U'_{x\mu}=\exp(i\lambda_{x+\mu})U_{x\mu}\exp(-i\lambda_x),
\label{u1}
\ee
where $\lambda_x$ is a site-dependent real variable.

Let us first consider the pure gauge system described by the energy $F_A$
alone.
In the case in which fluctuations of the vector potential $A_{x\mu}^{\rm em}$
are small, the above two versions belong to the same universality class,
i.e., the system is in the Coulomb phase. This is
because $F_A$ of (\ref{FA1}) approaches to Eq.(\ref{FA2}) due to the relation,
\be
&&U_{x\nu}^\ast U_{x+\nu,\mu}^\ast U_{x+\mu,\nu}U_{x\mu}+{\rm c.c.}=
2\cos\theta_{x\mu\nu},\nn 
&&\theta_{x\mu\nu}\equiv\nabla_\mu A_{x\nu}^{\rm em}-\nabla_\nu A_{x\mu}^{\rm em},\nn
&&\cos\theta_{x\mu\nu}\simeq 1-\frac{1}{2}\theta_{x\mu\nu}^2\ 
{\rm for\ small}\ \theta_{x\mu\nu}. 
\ee
For large fluctuations of vector potentials, 
the compact version generally allows 
topologically nontrivial excitations like
magnetic monopoles and may exhibit another phase called the confinement phase,
which is not allowed in the noncompact version. 

Next we consider the case in which the magnetic field is switched off   
by setting $A_{x\mu}^{\rm em}=0$. Then  the system (\ref{LGT1}) is reduced 
to the $|\phi|^4$ theory.
In the 3D $|\phi|^4$ theory, 
there exists a second-order phase transition
accompanying the spontaneous symmetry breaking of the 
global U(1) symmetry under the phase rotation, 
$\psi_x\to\exp(i\theta)\psi_x$. This broken phase 
is called Higgs phase and corresponds to the SC phase.
In the limit of $\lambda\to\infty$
with the ratio $\sigma/\lambda$ kept fixed to a finite negative value,
the system reduces to the so called XY model defined with 
$|\psi_x|=\sqrt{-\sigma/2\lambda}$, 
which is well known to exhibit a second-order transition as $K$ is varied. 
This limit corresponds to the London limit of the SC (or superfluidity), 
and the
phase transition in the $|\phi|^4$ theory for large $\lambda$ and 
that of the XY model belong to the same universality class. Even in this 
simplified model, the detailed critical behavior at the phase transition is 
{\em different from} that described by the mean-field theory (MFT). 
 
Finally, let us turn on the vector potential $A_{x\mu}^{\rm em}$.
In the continuum, it is shown\cite{coleman} 
that the second-order  transition for $\vec{A}^{\rm em}=0$ is changed 
to a first-order one as the  GL parameter 
$\kappa \propto \lambda/e^2$ is decreased.
On the lattice, the compact version of the system is 
studied in Ref.\cite{kajantie} 
and it is verified that the phase transition takes place 
as one varies $K$ with fixed $\kappa$.
More precisely, the phase transition
is of first order for small $\kappa$ and becomes second order
for large $\kappa$ as in the model defined in the continuum. 
Here we should note that the study in Ref.\cite{kajantie2}
shows that the critical behavior of the second-order transition
near the London limit is {\it not} in the same universality class 
as the 3D XY model due to the gauge-field fluctuations.

The noncompact lattice version is also studied in 
the London limit in Ref.\cite{chavel}. 
The corresponding energy is 
obtained from Eq.(\ref{LGT1}) by making the replacement,
\begin{eqnarray}
&&\frac{K}{2}\sum_{x,\mu}\left(\psi^\ast_{x+\mu}U_{x\mu}
\psi_x+{\rm c.c.}\right)
\nonumber \\
 && \hspace{1.5cm}
\rightarrow 
\frac{\tilde{K}}{2}\sum_{x,\mu}\left(e^{-i\varphi_{x+\mu}}
U_{x\mu}e^{i\varphi_x}+{\rm c.c.}\right),
\label{LGT2}
\end{eqnarray}
where $\varphi_x$ is the {\em phase} of the Cooper-pair (Higgs) field $\psi_x$,
and neglecting $F_\psi$ in Eq.(\ref{Fpsi}) because it becomes a constant.
In Ref.\cite{chavel} this U(1) gauge-Higgs model on the 
four dimensional lattice  is studied for large $1/e^2$,
which exhibits a second-order transition as $\tilde{K}$ 
is varied. This is consistent with the fact that, 
in the limit of $1/e^2\to\infty$, the system reduces to the 
four-dimensional XY model. 
This phase transition is induced by the condensation of vortex excitations
in $\varphi_x$. 
In the terminology of XY spin model, the XY spin
$\vec{\cal S}_x\equiv (\cos \varphi_x, \sin \varphi_x)^t$ 
has a definite amplitude $|\vec{\cal S}_x|=1$,
whereas the disorder phase $\la \vec{\cal S}_x\ra=0$ is possible due to 
the strong fluctuations of its angle $\varphi_x$. 
 
Then it is useful to rewrite the system in terms of topological excitations
such as vortices and monopoles.
In the London limit  of Eq. (\ref{LGT1}), 
the duality transformation can be perfomed\cite{chavel,es}, and
the free energy is expressed by the integer-valued vortex line-element 
variables $J_{\bar{x}\mu}$ and the integer-valued 
monopole-density varaiables $Q_{\bar{x}}\equiv \nabla_\mu J_{\bar{x}\mu}$ 
sitting on the dual lattice ($\bar{x}$ denotes its site) as
\be
F_v&=& 4\pi^2 K \sum_{\bar{x},\bar{y}}\left(\sum_{\mu}J_{\bar{x}\mu}
J_{\bar{y}\mu}
+\frac{1}{m_0^2}Q_{\bar{x}}Q_{\bar{y}}\right)
D_{\bar{x},\bar{y}}, \nn
D_{\bar{x},\bar{y}} &\simeq& 
\frac{\exp(-m_0r)}{m_0r},\ r=|\bar{x}-\bar{y}|,\ m_0^2=2Ke^2, 
\label{F_vortex}
\ee
where $D_{\bar{x},\bar{y}}$
is the 3D lattice Green's function with mass $m$. 
As explained in introduction, there appear no singularities in 
$F_v$ (\ref{F_vortex}).

For the noncompact version of the gauge system, 
it is shown in Ref.\cite{chavel} 
that no monopoles exist $Q_{\bar{x}}=0$ and 
only the closed vortex loops that satisfy 
$\sum_\mu \nabla_\mu J_{\bar{x}\mu}=0$ 
are allowed as expected. 
In the Coulomb phase with lower $K$ 
these vortex loops condense while in the Higgs phase 
with higher $K$ vortex loops are suppressed. 

On the other hand, for the compact version\cite{es},
open vortex strings may appear and a monopole 
should locate at every end of an open string such that 
$\sum_\mu \nabla_\mu J_{\bar{x}\mu}=Q_{\bar{x}} \neq 0$. 
Condensation of these monopoles
may imply sufficiently large fluctuations of 
$\vec{A}^{\rm em}$ and drive the system into the confinement 
phase\cite{confinement}. In fact, in the 3D compact case,
the system is known to stay always in the confinement 
phase\cite{confinement,janke}. In the 4D compact case, there is a 
gauge transition from the confinement phase to  the Coulomb phase as $1/e^2$ 
increases and a Higgs transition
from the Coulomp phase to the Higgs phase as $K'$ increases\cite{4du1higgs}.
Here let us comment on the 3D multi-component Higgs model in the 
compact version with $N$ Higgs fields in the London limit\cite{multihiggs}.
In contrast to the above 3D model with $N=1$, the model
with $N \geq 2$ supports the Higgs phase 
due to the extra phase degrees of freedom that are free from
coupling to the gauge field.

The above discussion clearly shows that the SC phase transition do take  
place even in the London limit for the gauge Higgs model 
in the noncompact gauge version with $N=1$
and in the compact version with $N \leq 2$, 
and topological excitations of SC order parameters, vortices,
play an important role for that.

Usually, the genuine transition temperature $T_c$ 
of the system (\ref{sGL}) is lower than  
the bare critical temperature $T^0_c$ due to fluctuation effect.
The radial degrees of freedom of $\psi_x$ 
may certainly contribute such renormalization of $T_c$, but 
should not change the universality class
of the continous phase transition we are to study 
because their fluctuations are massive. 

In the rest of the present paper, we shall study the FMSC state
by starting with a lattice model corresponding to
Eq.(\ref{LGT1}) in the London limit, in which
vortices are expected to be generated spontaneously, and therefore
nonperturbative study is indispensable for the investigation.

\subsection{GL thory in the continuum}
In the proposed GL theory\cite{MO,mineev}  for the FMSC materials
in the 3D continuum space at finite $T$,  the
free energy density 
$f_{\rm GL}$ for the SC state measured from the normal state is given as
\be
f_{\rm GL}&=&K\sum_{\mu}(D_\mu\vec{\psi})^\ast\cdot (D_\mu\vec{\psi})+
\alpha(T-T_{\rm{\tiny SC}}^0)\vec{\psi}^\ast\cdot\vec{\psi}\nn
&&+\lambda(\vec{\psi}^\ast\cdot\vec{\psi})^2
+K'\sum_{\mu}(\partial_\mu \vec{m})^2+(T-T_{\rm{\tiny FM}}^0)\vec{m}^2\nn
&&+\alpha_f(\vec{m}^2)^2+f_{\rm Z},\nn
D_\mu&=&\partial_\mu-2ieA_\mu,\ f_{\rm Z}=-J \vec{m}\cdot\vec{S},\nn
\vec{S}&=&i\vec{\psi}^\ast\times \vec{\psi},
\label{GL}
\ee
with the spatial direction index $\mu=1,2,3$. 
The three-component complex field $\vec{\psi}=
(\psi_{1},\psi_{2},\psi_{3})^{\rm t}$ 
is the spin-triplet SC order parameter, i.e., the Cooper-pair field
(we omit the spatial coordinate $x$ in the field $\vec{\psi}(x)$, etc.).
$\vec{\psi}$ is proprotional to the $\vec{d}$-vector in the
spin space,  $\vec{\psi}\propto \vec{d}$, 
and also the wave function of the $p$-wave
SC state in the real space as a result of the spin-orbit coupling.
In terms of $\vec{\psi}$, the magnetization (spin and angular momentum) 
$\vec{S}$ of  Cooper pairs  is expressed as in Eq.(\ref{GL})\cite{MO}.
The FM order is described by the magnetization field $\vec{m}$
of electrons that {\it do not} participate in the SC state.
$T_{\rm SC}^0$ and $T_{\rm FM}^0$ are the {\em bare} critical 
temperatures of SC and FM transitions, respectively.
Because $\vec{m}$ satisfies $\vec{\nabla}\cdot\vec{m}=0$,
it can be expressed in terms of the vector
potential $\vec{A}$ as $\vec{m}={\rm rot}\vec{A}$.
Because the Cooper-pair field bears the electric
charge $-2e$, it couples with $\vec{A}$ minimally via the
covariant derivative $D_\mu$ reflecting the 
electromagnetic gauge invariance.
We note that this vector potential $\vec{A}$ is {\it not}
the one for the external electro-magnetic field  
$A_{x\mu}^{\rm em}$ in Eq.(\ref{LGT1}).
$f_{\rm Z}$ is the Zeeman coupling term between 
$\vec{S}$ and $\vec{m}$. It  may induce the FMSC state\cite{MO},
in which $\langle\vec{m}\rangle\neq 0, \langle\vec{S}\rangle\neq 0$. 

The GL theory (\ref{GL}) and the related ones have been studied 
so far by means of MFT-type methods\cite{GL}.
However, the gauge coupling between $\vec{A}$ and $\vec{\psi}$ 
makes a simple MF approximation assuming, e.g.,
a constant SC order parameter and ignoring topologically
nontrivial fluctuations unreliable for the description of the FMSC state.
This is also indicated as we explained 
for the model of $s$-wave SC state, Eqs.(\ref{sGL}) and (\ref{LGT1}).
Therefore $f_{\rm GL}$ in Eq.(\ref{GL}) is a kind of {\em frustrated system
of the AF and SC}. We dare to use the word ``frustrated'' here.
This is because the case with
a finite magnetization $\vec{m}\neq 0$ corresponds to the case
with a non-vanishing external magnetic field; a well known case
of frustration. In fact, the phase of matter field there should
acquires a finite additional phase when it is rotated along a closed loop 
and so the phase is not determined uniquely except for an integer magnetic 
flux inside the loop.   
In other words, vortices are to be generated because of the presence of
the nonvanishing magnetization.

It is of course important to study the set of GL equations 
derived from $f_{\rm GL}$ in Eq.(\ref{GL}).
The GL equations should be solved self-consistently to obtain $\vec{\psi}(x)$
and $A_\mu(x)$.
In the FM phase, $\vec{\psi}(x)$ describes  multi-vortex states
and the vector potential $A_\mu(x)$ is determined by the 
locations of vortices in addition to the other terms in $f_{\rm GL}$
including the magnetization $\vec{m}(x)$.

In the present paper, we shall introduce a GL theory defined
on a cubic lattice which is a discretized version of Eq.(\ref{GL}), 
and study its phase structure etc by means of the Monte-Carlo simulations.
In this approach, all relevant fluctuations of $\vec{\psi}$
and $A_\mu$ are taken into account. 
Some related lattice model describing a two-component SC was studied
and interesting results were obtained\cite{TCSC}.
The above two  approaches, MFT of the GL solutions and the numerical study
of the GL theory, are complementary and not exclusive each other.

\subsection{Derivation of the lattice model}
As announced in Sect.I,  we introduce an effective 
lattice gauge model based on the GL theory
in the continuum (\ref{GL}) by making a couple of simplifications.
We stress that topological defects
such as  vortices are allowed on the lattice
without introducing an additional short-distance cutoff for vortex cores.
This point is quite important because it is expected that such nontrivial
excitations are generated in the FMSC phase.

As the first step of simplification, we consider  
the ``London limit" of $\vec{\psi}$ such that
$\vec{\psi}^\ast\cdot\vec{\psi}=$ const., neglecting its radial
fluctuations.
As discussed in Sect.II.A for the $s$-wave SC model, 
this is legitimate because 
the phase degrees of freedom of $\vec{\psi}$
play an essential role for (in)stability of the SC state.

Second, the FMSC materials  have a FM easy-axis, which we choose
the $z$-direction ($\mu=3$)\cite{anisotoropy}.
The Zeeman coupling $f_{\rm Z}$ prefers such that
$\psi_{\uparrow\uparrow}\propto \psi_1+i\psi_2$
or $\psi_{\downarrow\downarrow}\propto \psi_1-i\psi_2$ takes large amplitude.
In fact, as the Fermi surfaces of up and down-spin electrons have different 
energies due to  $f_{\rm Z}$, 
the Cooper-pair amplitude of mixed spins, 
$\psi_3=\psi_{\uparrow\downarrow}$, is  small.

Therefore, in the effective model, we ignore $\psi_3$ as in the
previous studies\cite{MO,Harada,anisotoropy}, 
and consider only the two-component complex field  
$(\psi_1,\psi_2)^{\rm t}$ in the London limit,
\be
&& (\psi_1,\psi_2,\psi_3)\to (\psi_1,\psi_2,0)  \nn
&& {\rm with} \ |\psi_1|^2+|\psi_2|^2= 
{\alpha \over 2\lambda}(T^0_{\rm {\tiny SC}}-T).
\label{psitoz}
\ee
Here we introduce a two-component complex field $z$ 
that is the normalized Cooper-pair field;
$z=(z_{1},z_{2})^{\rm t}$ satisfying 
\be
|z_{1}|^2+|z_{2}|^2=1,
\label{CP1}
\ee 
and from (\ref{psitoz})
\be
(\psi_1, \psi_2)^{\rm t}=
\sqrt{{\alpha \over 2\lambda}(T^0_{\rm {\tiny SC}}-T)} \;
(z_1, z_2)^{\rm t}.
\ee
We note that two-component complex variables 
satisfying Eq.(\ref{CP1}) such as $z(x)$ is called a CP$^1$ 
(complex projective) field.
In term of $z(x)$, the first term of $f_{\rm GL}$ in Eq.(\ref{GL})
is given as 
\be
{K\alpha \over 2\lambda}(T^0_{\rm {\tiny SC}}-T)
\sum_{\mu}(D_\mu z)^\ast\cdot (D_\mu z).
\label{fGLz}
\ee

 Now let us consider the effective lattice model on the 3D cubic 
lattice. Its GL free-energy density $f_x$ at the site $x$  is given 
up to an irrelevant constant by
\begin{eqnarray}
f_x &=& -\frac{c_1}{2}\sum_{\mu=1}^3\sum_{a=1}^2\left( 
\bar{z}_{x+\mu,a}U_{x\mu}z_{xa}+ {\rm c.c.}\right)-
c_2 \vec{m}_x^2\nn
&&-c_3 \vec{m}_x\cdot\vec{S}_x +c_4(\vec{m}_x^2)^2  
-c_5  \sum_{\mu}\vec{m}_{x}\cdot\vec{m}_{x+\mu},\nn
U_{x\mu} &\equiv& \exp(iA_{x\mu}).
\label{FLGL}
\end{eqnarray}
The five coefficients $c_i$ $(i=1\sim 5)$ in (\ref{FLGL})
 are real nonnegative parameters 
that are expected to distinguish various materials in various environments.
$z_x =(z_{x1}, z_{x2})^{\rm t}\ (\sum_{a=}|z_{xa}|^2=1)$
 is the  CP$^1$ variable at the site $x$ and plays the role of SC
 order-parameter field. 
$U_{x\mu}$ is the exponentiated vector 
potential\cite{2e}
put on the link ($x,x+\mu$). 
$\vec{m}_x=(m_{x1},m_{x2},m_{x3})^{\rm t}$ is 
the magnetic field made out of $A_{x\mu}$ as
\begin{eqnarray}
\hspace{-0.8cm}
m_{x\mu} \equiv\! \sum_{\nu,\lambda=1}^3
 \epsilon_{\mu\nu\lambda}\nabla_\nu A_{x\lambda}, \
 \nabla_\nu A_{x\lambda} \equiv A_{x+\nu,\lambda}\!-\!A_{x\lambda}.
\label{rota}
\end{eqnarray}
$\vec{m}_x$ serves as the FM order-parameter field. 
$\vec{S}_x = (0,0,S_{x3})^t$ is  an
Ising-type spin vector of Cooper pairs made out of $z_{xa}$ as
\be
S_{x3} \equiv i (z_{x1}^* z_{x2}-z_{x2}^* z_{x1})
\propto |\psi_{\uparrow\uparrow}|^2-|
\psi_{\downarrow\downarrow}|^2.
\ee
As for the case of (\ref{u1}), 
$f_x$ is invariant under a U(1) gauge transformation,
\be
z_{xa} &\rightarrow& z_{xa}'=\exp(i\lambda_x) z_{xa},\nn
U_{x\mu}&\rightarrow& U_{x\mu}'=
\exp(i\lambda_{x+\mu})U_{x\mu}\exp(-i\lambda_x).
\label{gaugesym}
\ee
We note that both $\vec{m}_x$ and $\vec{S}_x$ are gauge invariant.

The meaning of each term in $f_{x}$ is as follows.
The $c_1$-term describes a hopping of Cooper pairs.
From (\ref{fGLz}), it is obvious that
\be
c_1 \sim K\alpha \lambda^{-1}(T^0_{\rm {\tiny SC}}-T)a,
\label{c1}
\ee
where $a$ is the lattice spacing.
At sufficiently large $c_1$, the $c_1$-term may stabilize
the phase of $z_{xa}$, and then 
a {\em coherent condensation of the phase degrees of freedom of}  
$z_x$ {\it induces the superconductivity}.
The $c_2$ and $c_4$-terms are the quartic GL potential of 
$\vec{m}_x$ and favor a finite amount of {\it local} magnetization
$\la \vec{m}_x \ra \neq 0$ (note that we take $c_2 > 0$). 
Again we note these terms controls intrinsic magnetrization
and different from $F_A$ of (\ref{LGT1}) for the fluctuationg but 
external magnetic field.
The $c_5$-term enhances uniform configurations of 
$\vec{m}_x$, i.e.,  a FM long-range order signaled by
a finite magnetization,
$\lim_{|x-x'|\to \infty}\la\vec{m}_x\cdot\vec{m}_{x'}\ra \neq 0$.
The $c_3$-term is the Zeeman coupling, which favors 
collinear configurations of $\vec{m}_x$ and $\vec{S}_x$, 
namely the coexistence of ferromagnetism and superconductivity.

The partition function $Z$ at $T$ is given by
the integral over a set of two fundamental fields $z_{xa}$ and  $A_{x\mu}$ as
\be
Z &=& \int[dz][dA]\exp(-\beta F),\ \beta=T^{-1},\ F=\sum_x f_x,\nn 
\left[dz\right]&=&\prod_x d^2z_{x1}d^2z_{x2}\ \delta(|z_{x1}|^2+|z_{x2}|^2-1),
\nn
\left[dA\right]&=&\prod_{x,\mu} dA_{x\mu},\ A_{x\mu} \in (-\infty,\infty). 
\label{Z}
\ee

The coefficients $c_i$ in $f_x$ may have nontrivial
$T$-dependence as Eqs.(\ref{GL}) and (\ref{c1}) suggest. However,
in the present study we consider the response of the system 
by varying the ``temperature" $T \equiv 1/\beta$
defined by $\beta$,  
an overall prefactor in Eq.(\ref{Z}), while keeping $c_i$ fixed.
This method corresponds to well-known studies such as the FM transition
by means of the $O(3)$ nonlinear-$\sigma$ model\cite{sigma} and 
the lattice gauge-Higgs models  discussed in Sect.II.A, and sufficient to
determine the critical temperature.
See later discussion leading to Eq.(\ref{TTSC2}).

\section{Numerical studies}
\setcounter{equation}{0}
\subsection{FM ands SC phase transition and Meissner effect}

For explicit procedures of our Monte-Carlo simulations, we 
first prepare a 3D lattice of the size of
$(2+L+2)^2\times L$, namely the lattice coordinates
runnning as $x_1, x_2 = 0,...,L+3, x_3 = 0,...,L-1$. 
The extra width 2+2 in the $\mu=1,2$ directions is introduced
as a buffer zone in which the suppercurrrent is damped.
The calculations of physical quantities are done in the central
region $R$ of the size $L^3$ to suppress the effects of the boundary.
For the boundary condition we choose the periodic 
boundary condition in the $\mu=3$ direction.
For the $\mu=1,2$ directions, we first note that
the supercurrent density $j^{\rm SC}_{x\mu}$  on the lattice 
is expressed as
\be
j^{\rm SC}_{x\mu} \propto {\rm Im}\ (\bar{z}_{x+\mu}U_{x\mu}z_x).
\ee
Then we impose
\be
\hspace{-0.3cm}
z_{x+\mu} - U_{x\mu}z_x = 0\
 {\rm for}\ &&
 x=(0,x_2,x_3), \mu=1,\nn
&&x=(L+2,x_2,x_3), \mu=1, \nn
&&x=(x_1,0,x_3), \mu=2, \nn
&&x=(x_1,L+2,x_3), \mu=2,
\label{bc1}
\ee
whereas we impose the free boundary condition on $A_{x\mu}$.

The above condition (\ref{bc1}) is gauge invariant and assures us
that $j^{\rm SC}_{x\mu} =0\ (\mu=1,2)$ on the boundary surfaces
in the $\mu-3$ plane; the supercurrent do not leak out 
of the SC material put in the region $R$.
We note this boundary condition enhances the 
third-component of the FM order so that   
$\langle \vec{m}_x \rangle=(0,0,m_3)^{\rm t}$.
This fact is traced back to our choice $\psi_3=0$ in Eq.(\ref{psitoz}) 
and reflects the experimental fact that the real FMSC materials exhibit the 
Ising-type anisotoropy of the FM magnetization.
Note also that the conventional periodic boundary condition
for $A_{x\mu}$ in all the three directions implies 
$\sum_x\vec{m}_x = 0$ and the FM order cannot exist.

We use standard Metropolis algorithm for the lattice size
up to $L=20$. The typical sweeps for measurement is 
$(30000\sim 50000)\times 10$ and the acceptance ratio is $40\%\sim50\%$.

Explicitly, we calculate the internal energy $U$,
the specific heat $C$ of the central region $R$, 
the FM magnetization  $m_\mu$,
\be
U&=&\frac{1}{L^3}\langle F' \rangle,\
C=\frac{1}{L^3}\langle (F'-\langle F' \rangle)^2\rangle,\
F'= \sum_{x \in R} f_x,\nn
m_\mu&\equiv&\frac{1}{L^3}\langle |\sum_{x \in R} m_{x\mu}| \rangle,
\ee
and the normalized correlation functions, 
\be
G_{m}(x-x_0)&=&
\la\vec{m}_{x}\cdot\vec{m}_{x_0}\ra/\la\vec{m}_{x_0}\cdot\vec{m}_{x_0}\ra,\nn
G_{S}(x-x_0)&=&\la S_{x3}S_{x_0,3}\ra/\la S_{x_0,3}S_{x_0,3}\ra,
\ee
where $x_0$ is chosen on the boundary of $R$ such as $(3, 2+L/2, z)$.

\begin{figure}[b]
 \begin{minipage}{0.45\hsize}
  \begin{center}
  \hspace{-0.6cm}
   \includegraphics[width=4.3cm]{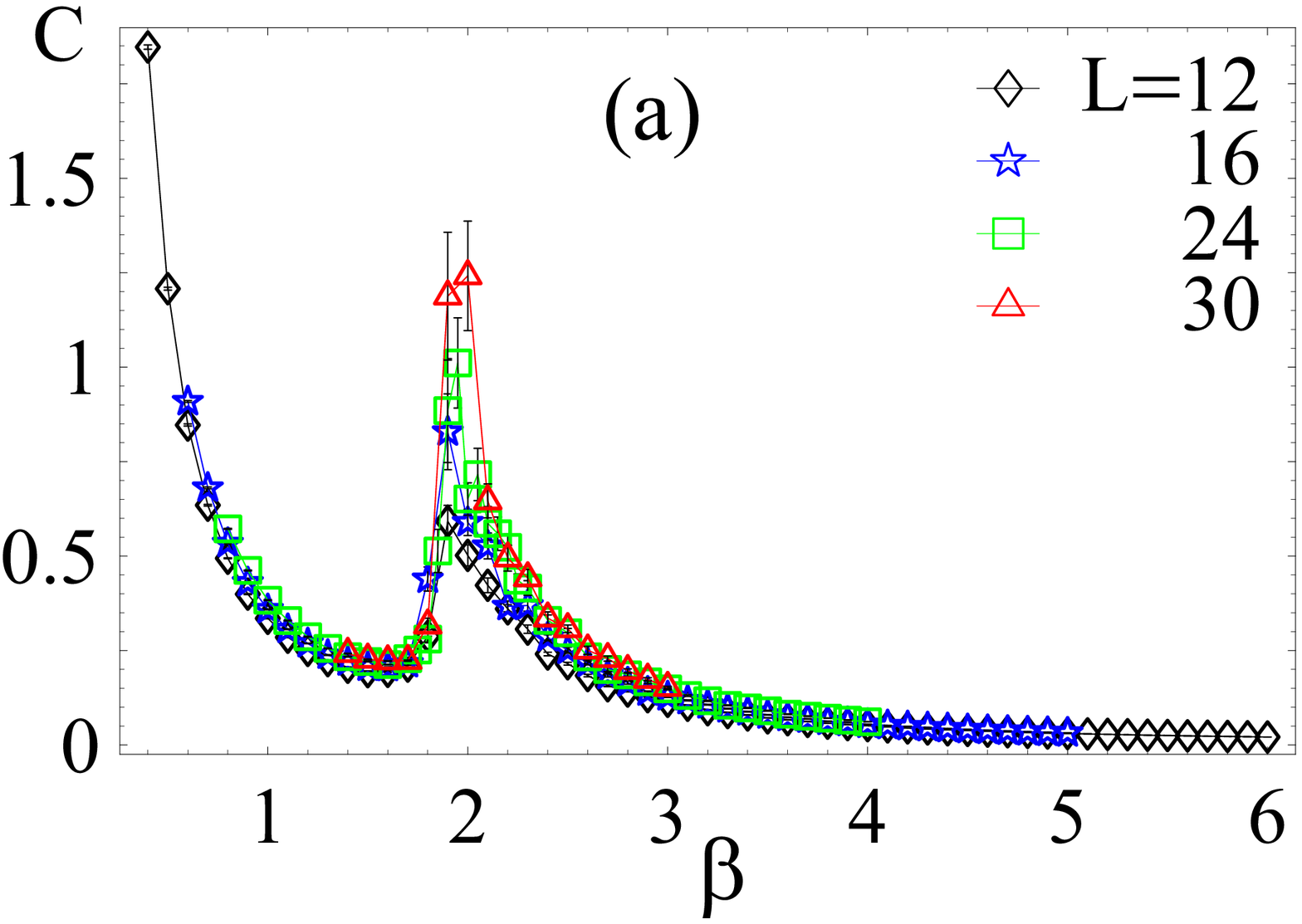}
  \end{center}
 \end{minipage}
     \hspace{-0.6cm}
 \begin{minipage}{0.45\hsize}
  \begin{center}
   \vspace{-0.6cm}
      \includegraphics[width=4.6cm]{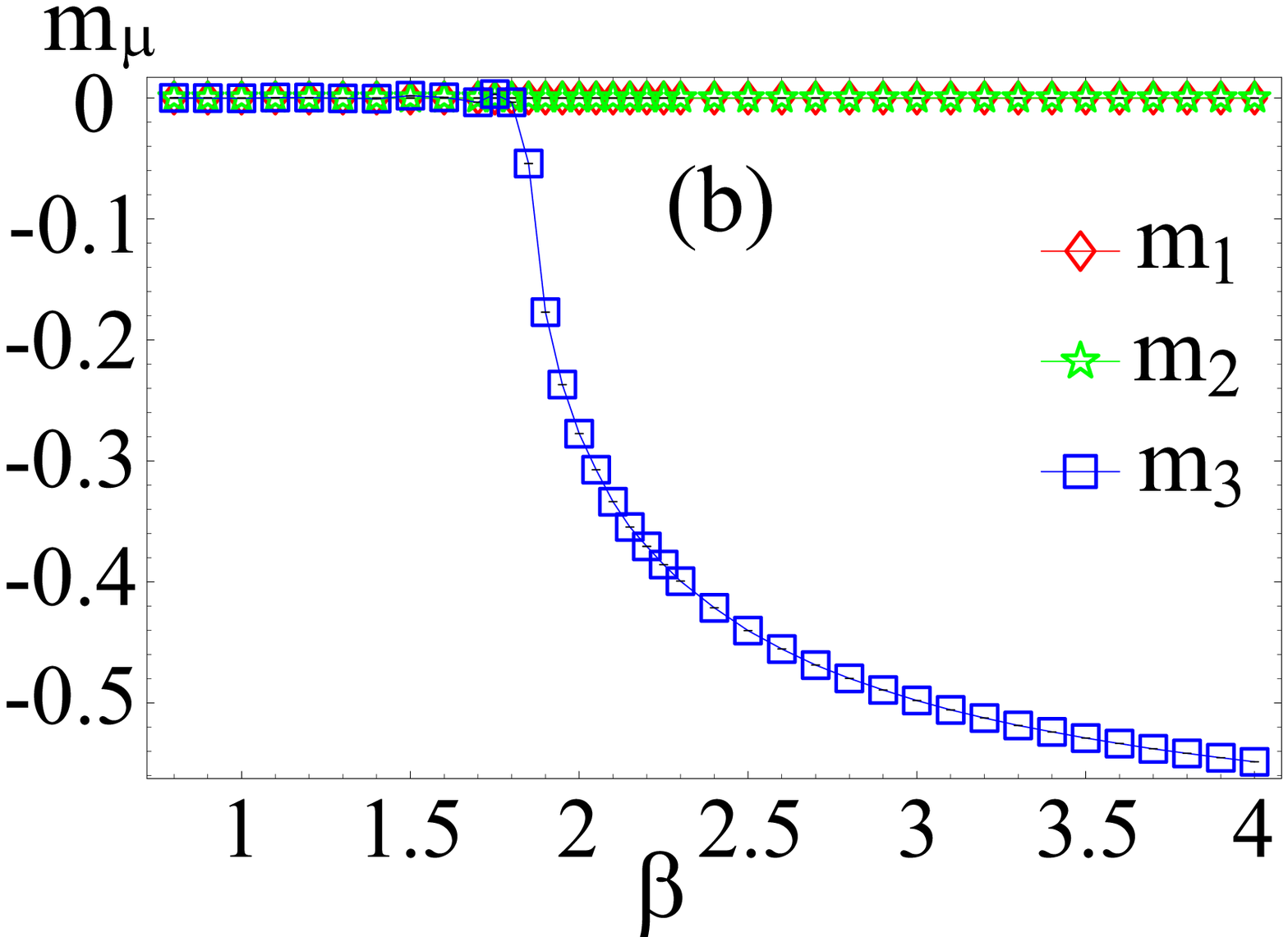}
  \end{center}
 \end{minipage}
\vspace{-0.2cm}
\caption{
(a) Specific heat  for
$(c_2,c_4,c_5)=(0.5,4.0,1.0)$ and $c_1=c_3=0$.
At $\beta\simeq 2.0$, $C$ exhibits a sharp peak indicating
a second-order FM phase transition.
(b) Each component of magnetization $m_\mu$
vs $\beta$.
For $T<T_{\rm FM}$, $m_3$ develops, whereas 
$m_1$ and $m_2$ are zero within the errors as expected.
}
\label{C:mag}
\end{figure}

We first show that the model (\ref{Z}) exhibits a FM phase transition
as $T$ is lowered.
To this end, we put $c_1=c_3=0$ and $(c_2,c_4,c_5)=(0.5,4.0,1.0)$,
and increase $\beta$ in the Boltzmann factor of Eq.(\ref{Z}). 
In Fig.\ref{C:mag}
we show $C$ and $m_\mu$. 
It is obvious that a second-order phase transition to the FM state takes place
at $\beta_{\rm FM}=1/T_{\rm FM}\simeq 2.0$. 
We verified that other cases with various values of $c_{2,4,5}$  
exhibit similar FM phase transitions.

\begin{figure}[t]
\vspace{-1cm}
\begin{center}
\hspace{-3cm}
\includegraphics[width=9cm]{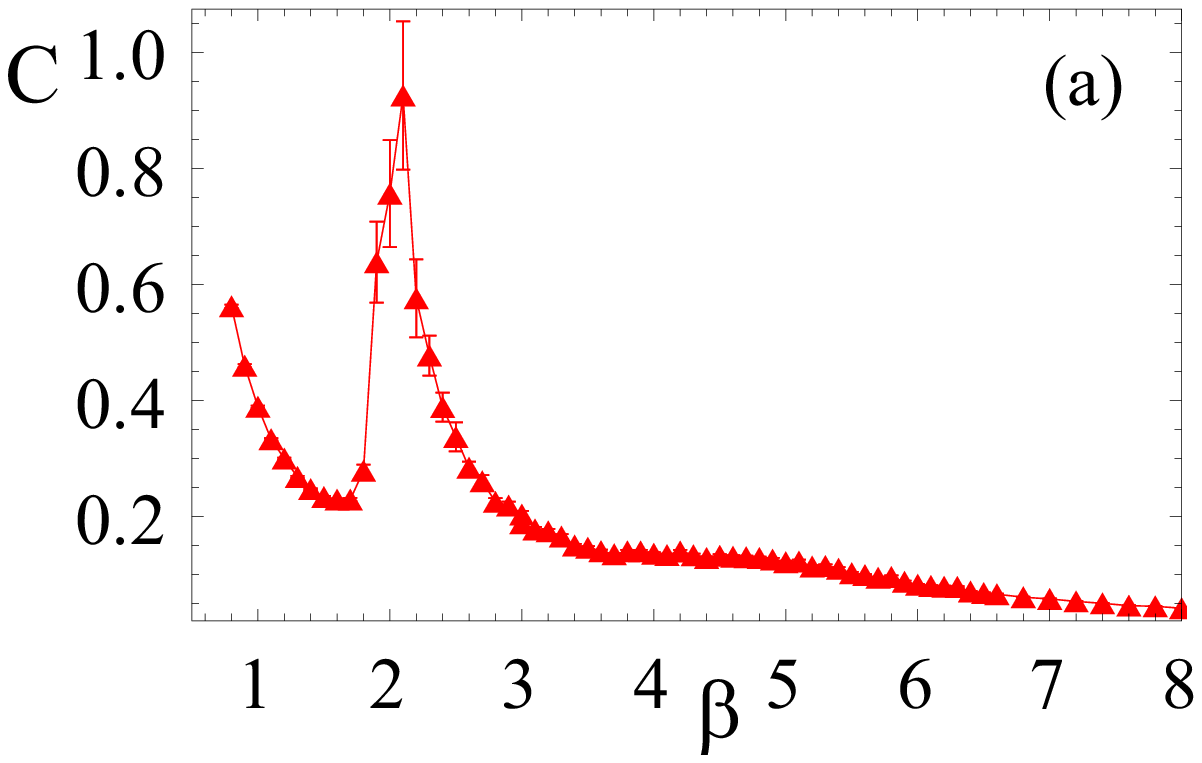}
\end{center}
\vspace{-0.5cm}
 \begin{minipage}{0.45\hsize}
  \begin{center}
  \hspace{-0.6cm}
\includegraphics[width=4.0cm]{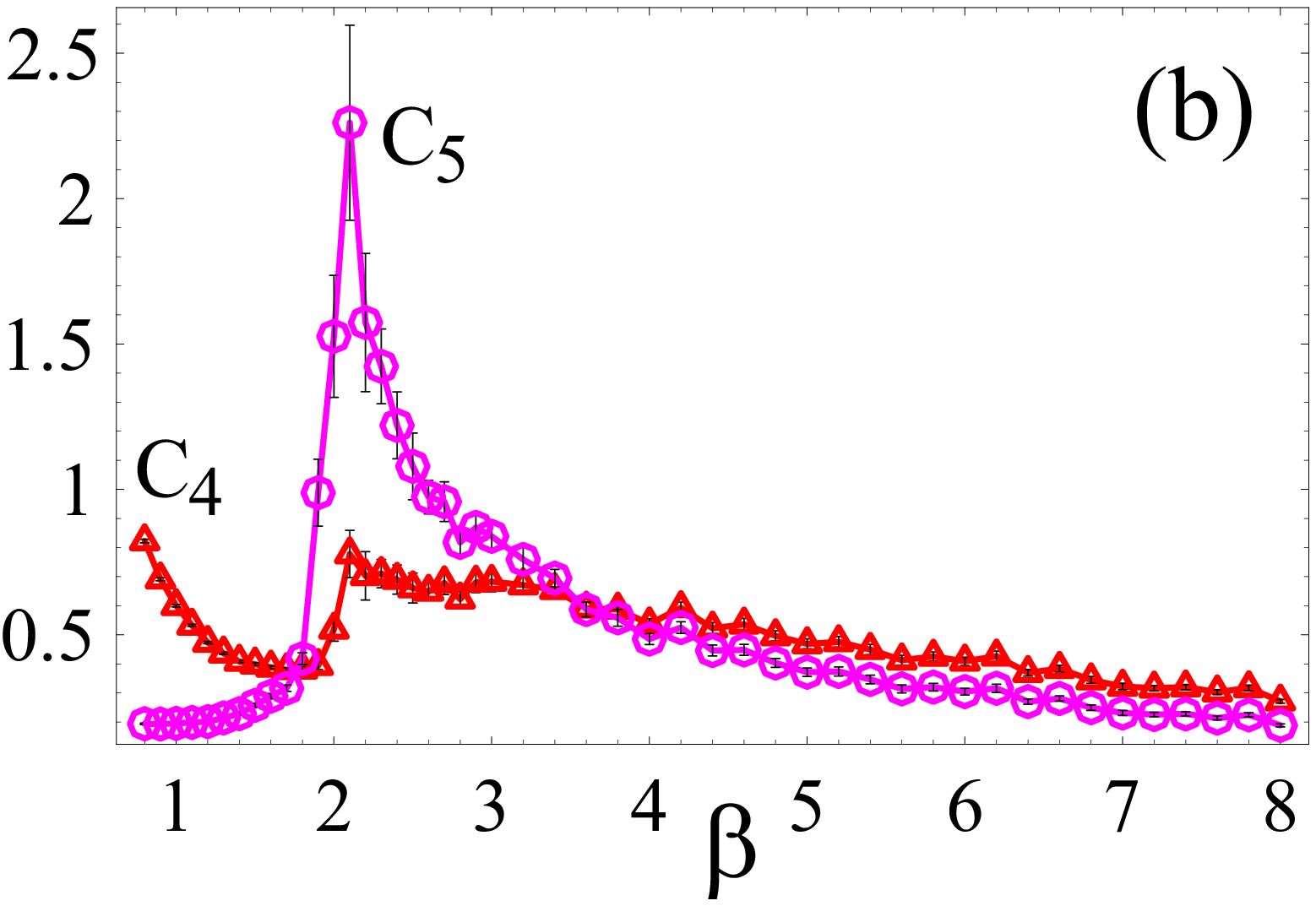}
  \end{center}
 \end{minipage}
     \hspace{-0.2cm}
 \begin{minipage}{0.45\hsize}
  \begin{center}
   \vspace{-0.9cm}
\includegraphics[width=4.2cm]{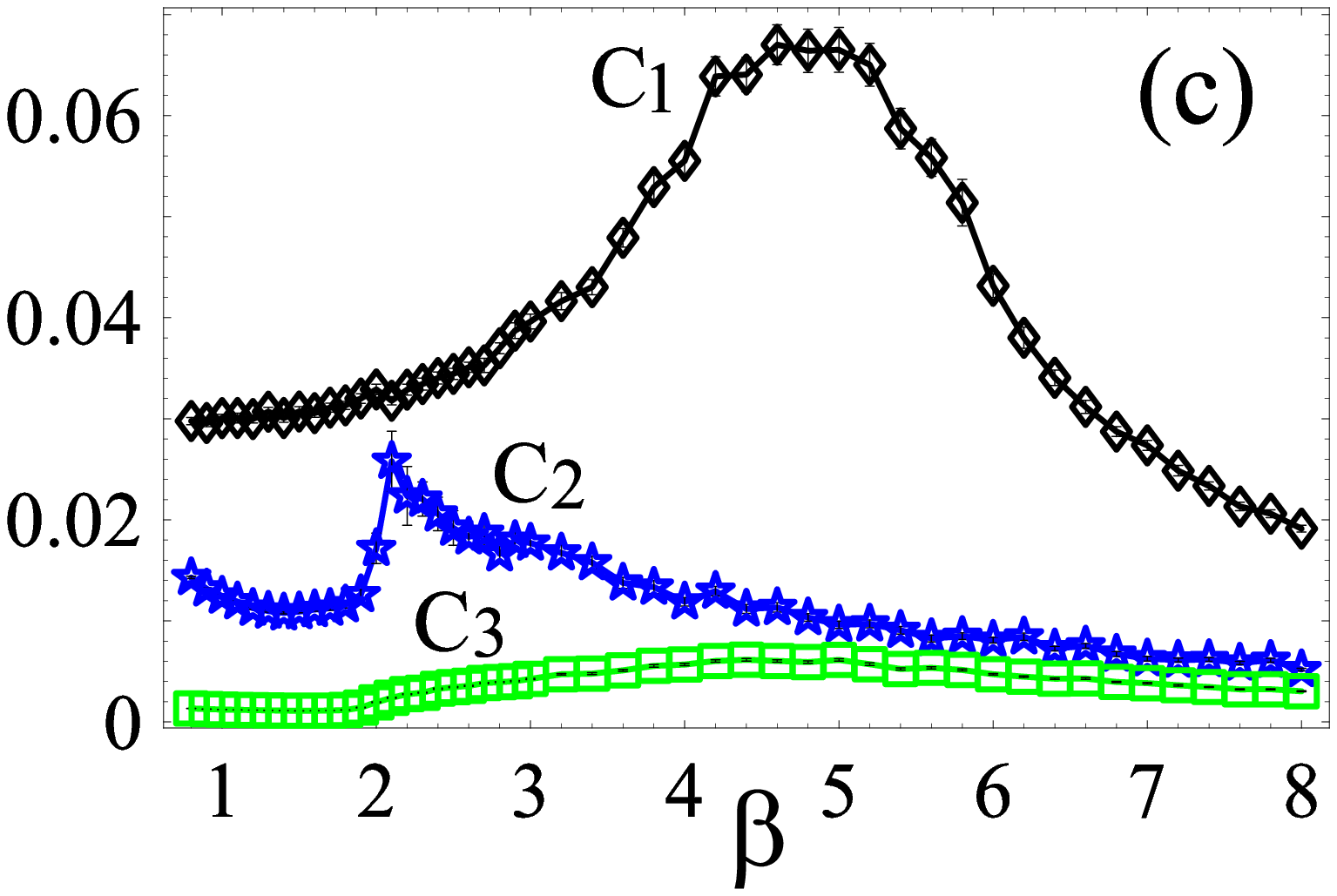}
  \end{center}
 \end{minipage}
 \vspace{-1.0cm}
\caption{
(a) Specific heat vs $\beta$ for $(c_1,c_3)=(0.2,0.2)$ and
$(c_2,c_4,c_5)=(0.5,4.0,1.0)$ ($L=20$). 
There are a large peak at $\beta \simeq 2.1$ and a small one at
$\beta \simeq 4.5$.
(b,c) Specific heat $C_i$ of each term of $F$ in (\ref{Z}) vs $\beta$.
The small and broad peak at $\beta \simeq 4.5$ in $C$ is related to
fluctuations of $c_1$-term. 
}
\label{C:magsc}
\end{figure}

Next, let us study the SC phase transition.
Here it is useful to consider the case of all $c_i=0$ except for $c_1$.
Then the model is related to the CP$^1$+U(1) lattice gauge 
theory\cite{takashima} which has the energy of the form 
$f_r = -(c_1/2)\sum(\bar z U z + {\rm c.c.})
+F_A$ where $F_A$ is the compact version (\ref{FA2}).
In fact, they agree by setting $1/e^2=0$.
The phase structure of this model is studied in Ref.\cite{takashima}
and it is found that the phase transition from the confinment phase 
to  the Higgs phase takes place at $c_1 \simeq 2.85$ for $1/e^2=0$.
Thus the SC state exists at sufficiently large $c_1$.

For simulation, we put $(c_1,c_3)=(0.2,0.2)$ and
$(c_2,c_4,c_5)=(0.5,4.0,1.0)$.
In Fig.\ref{C:magsc}a, we show $C$ vs $\beta$.
There are a large and sharp peak at $\beta \simeq 2.1$ and a small and 
broad one at $\beta \simeq 4.5$.
In order to understand physical meaning of the second peak,
it is useful to measure ``specific heat" of each term $f_i$ in the
free energy (\ref{FLGL}) defined by $C_i=\la(F'_i-\la F'_i\ra)^2\ra/L^3$.
Fig.\ref{C:magsc}c shows that the specific heat
of the $c_1$-term has a relatively large and broad peak at
$\beta \simeq 4.5$.
Then we conclude that the SC phase transition takes place
at $\beta_{\rm SC}\simeq 4.5$.

\begin{figure}[b]
\begin{center}
\hspace{-0.3cm}
\includegraphics[width=4.3cm]{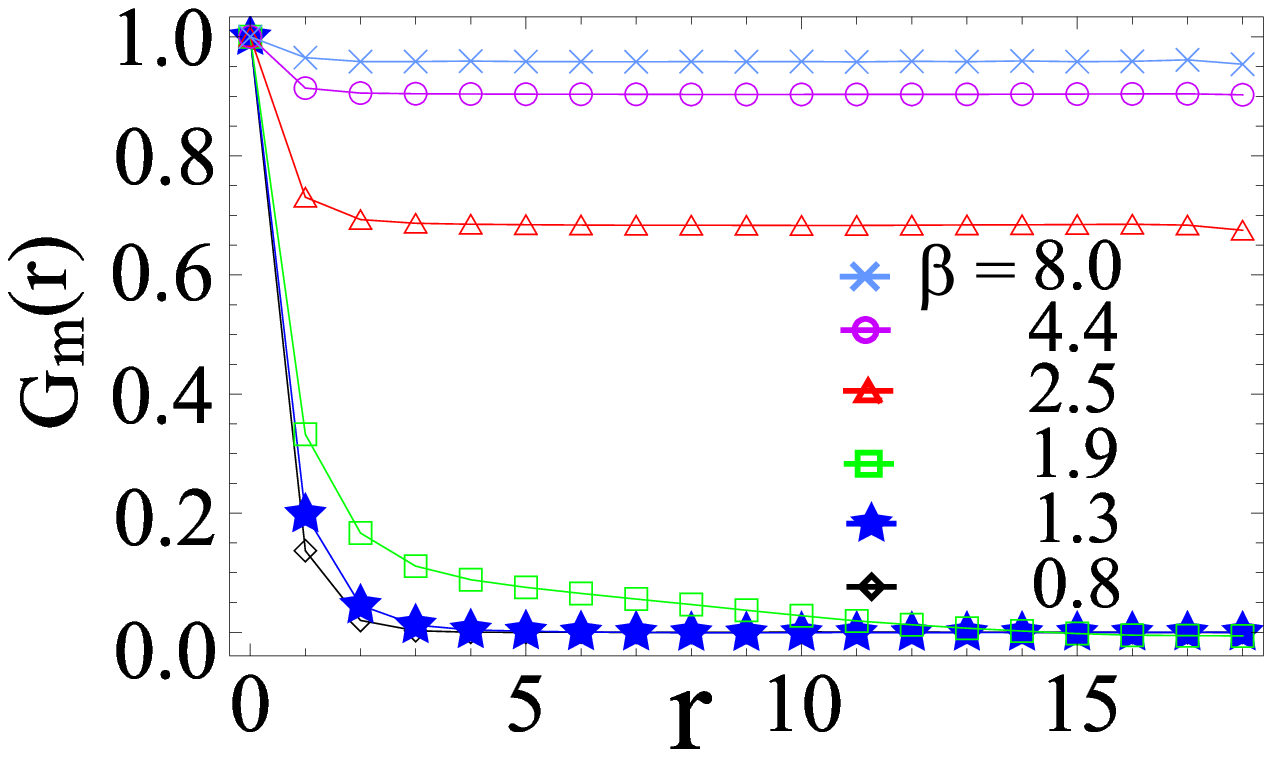}
\hspace{-0.0cm}
\includegraphics[width=4.3cm]{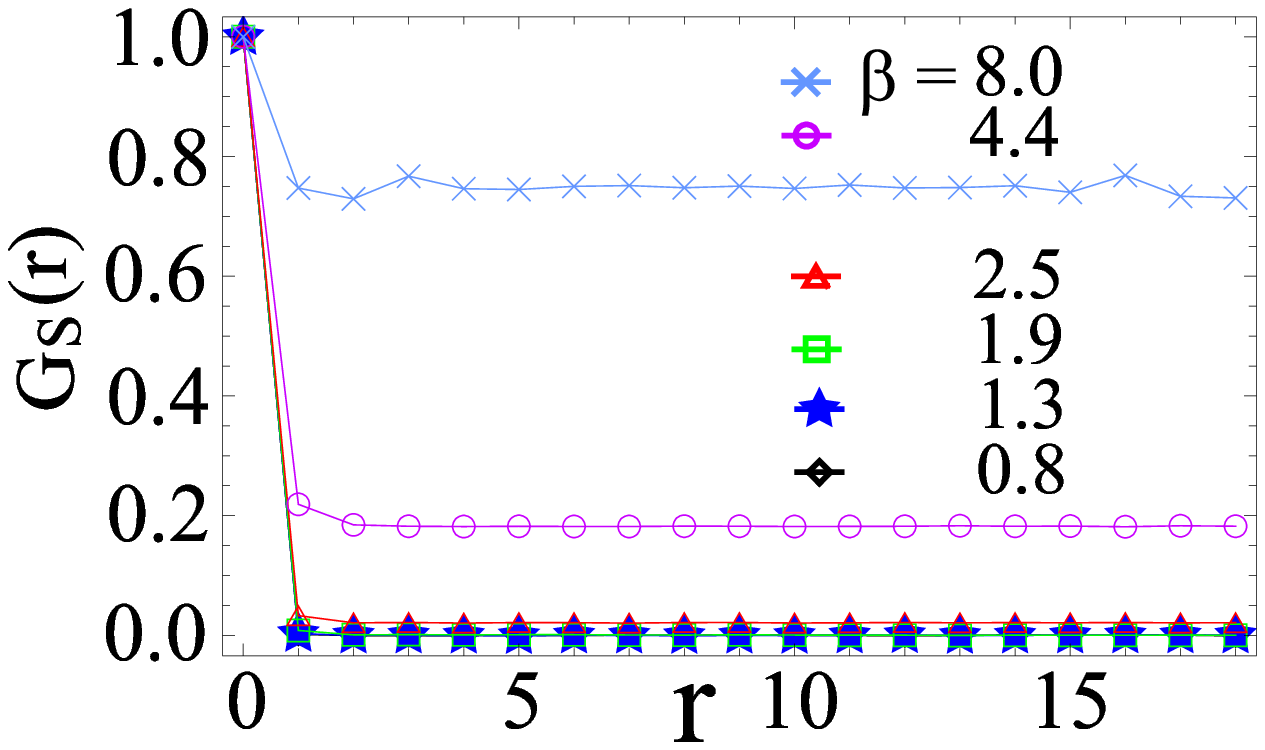}
\end{center}
\vspace{-0.3cm}
\caption{
Correlation functions $G_{m}(r)$ and $G_{S}(r)$ at various $T$'s
for $L=20$.
$c_i$'s are the same as in Fig.\ref{C:magsc}. 
}
\label{correlation1}
\end{figure}

To verify this conclusion, we show 
$G_{m}(r)$ and $G_{S}(r)$ in Fig.\ref{correlation1}.
At $\beta=2.5$, $G_{m}(r)$ exhibits a 
finite amount of the FM order, whereas $G_{S}(r)$
decreases very rapidly to vanish.
This means that, as $T$ is decreased, 
the FM transition takes place first and then
the SC transition does.
Therefore, for $\beta\geq\beta_{\rm SC}\simeq 4.5$, the FM and SC 
orders coexist.

It is interesting to clarify the relation between the bare transition 
temperature $T^0_{\rm SC}$ in Eq.(\ref{GL}) and the genuine 
transition temperature $T_{\rm SC}$.
From Eq.(\ref{Z}), any physical quantity is a function of $\beta c_i$.
In the numerical simulations, we fix the values of $c_i$ and vary $\beta$
as explained.
Then the result $\beta_{\rm SC} \simeq 4.5$ means
\be
\beta c_1|_{T=T_{\rm SC}}=4.5\times 0.2.
\label{TTSC}
\ee
By using Eq.(\ref{c1}), this gives the following relation;
\be
&&{1 \over T_{\rm SC}}{K\alpha (T^0_{\rm SC}-T_{\rm SC}) a\over \lambda}
=0.90,\nn
&&
T_{\rm SC}=\Big(1+{0.90\ \lambda \over K\alpha\ a}\Big)^{-1}T^0_{\rm SC}.
\label{TTSC2}
\ee
Eq.(\ref{TTSC2}) shows that the transition temperature
is lowered by the fluctuations of the phase degrees of freedom of
Cooper pairs.
We expect that a relevant contribution to lowering the SC transition 
temperature comes from vortices that are generated spontaneously 
in the FMSC as we show in Sec.III.B.

After having confirmed that the genuine critical temperature can be
calculated by the critical value of $\beta$ with fixed $c_i$, 
we use the word temperature in the rest of the paper just as the one defined 
by $T\equiv 1/\beta$ while $c_i$ are $T$-independent parameters.

\begin{figure}[b]
\begin{center}
\includegraphics[width=5cm]{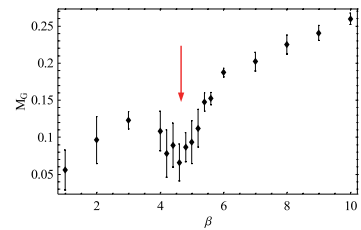}
\end{center}
\vspace{-0.7cm}
\caption{
Gauge-boson mass $M_{\rm G}$ of the external magnetic 
field propagating in $x$-$y$ plane vs $\beta$ for 
the same $c_i$ as in Fig.\ref{C:magsc}
and ${c_2}'=3.0$ ($L=16$).
At the SC phase transition point $\beta_{\rm SC}\simeq 4.5$
(indicated by an arrow) determined by the peak of $C$,
$M_{\rm G}$ starts to increase from small values. 
}
\label{MG}

\end{figure}

Meissner effect is one of the most important phenomena for a SC order.
To study it, we follow the following steps\cite{takashima};
(i) introduce a vector potential $A^{\rm ex}_{x\mu}$ for 
an external magnetic field,
(ii) couple it to Cooper pairs by replacing
$U_{x\mu} \to U_{x\mu}\exp(iA^{\rm ex}_{x\mu})$ in 
the $c_1$ term of $f_x$ 
and  add its magnetic term $f^{\rm ex}_{x}
=+{c_2}'(\vec{m}_x^{\rm ex})^2\ (c_2' > 0)$ to $f_{x}$ 
with $\vec{m}^{\rm ex}_r$  defined in the same way as (\ref{rota})
by using $A^{\rm ex}_{x\mu}$, 
(iii) let $A^{\rm ex}_{x\mu}$ fluctuate together with $z_{xa}$ and
$A_{x\mu}$ and   measure an effective mass $M_{\rm G}$
of $A^{\rm ex}_{x\mu}$ via the decay of correlation functions of
$\vec{m}_x^{\rm ex}$.
The result of $\vec{m}^{\rm ex}_x$ {\em propagating in the 1-2 plane} 
is shown in Fig.\ref{MG}.
It is obvious that the mass $M_{\rm G}$ starts to develop at
the SC phase transition point, and we conclude that Meissner effect
takes place in the SC state.

\subsection{SC transition and vortices in a constant magnetic field}

Because the observed SC state in Figs.\ref{C:magsc} and
\ref{correlation1} coexists with the FM order, 
it is expected that vortices of the SC order parameter 
are induced there spontaneouly\cite{vortex}.
To verify this expectation, we set the vector potential $A_{x\mu}$
to a position-dependent but nonfluctuating value  that corresponds
to a uniform magnetic field in the third-direction, 
and study the behavior of $z_{xa}$ itself.
In this case, the free energy $f_{x}$ loses the local
gauge symmetry (\ref{gaugesym}), 
and therefore the correlation function of $z_{xa}$,
\be
G_z(x-x_0)=\langle \bar{z}_{x}\cdot z_{x_0}\rangle,
\ee
has nonvanishing values in the SC phase.

\begin{figure}[t]
\begin{center}
\vspace{-0.3cm}
\hspace{-0.5cm}
\includegraphics[width=4.3cm]{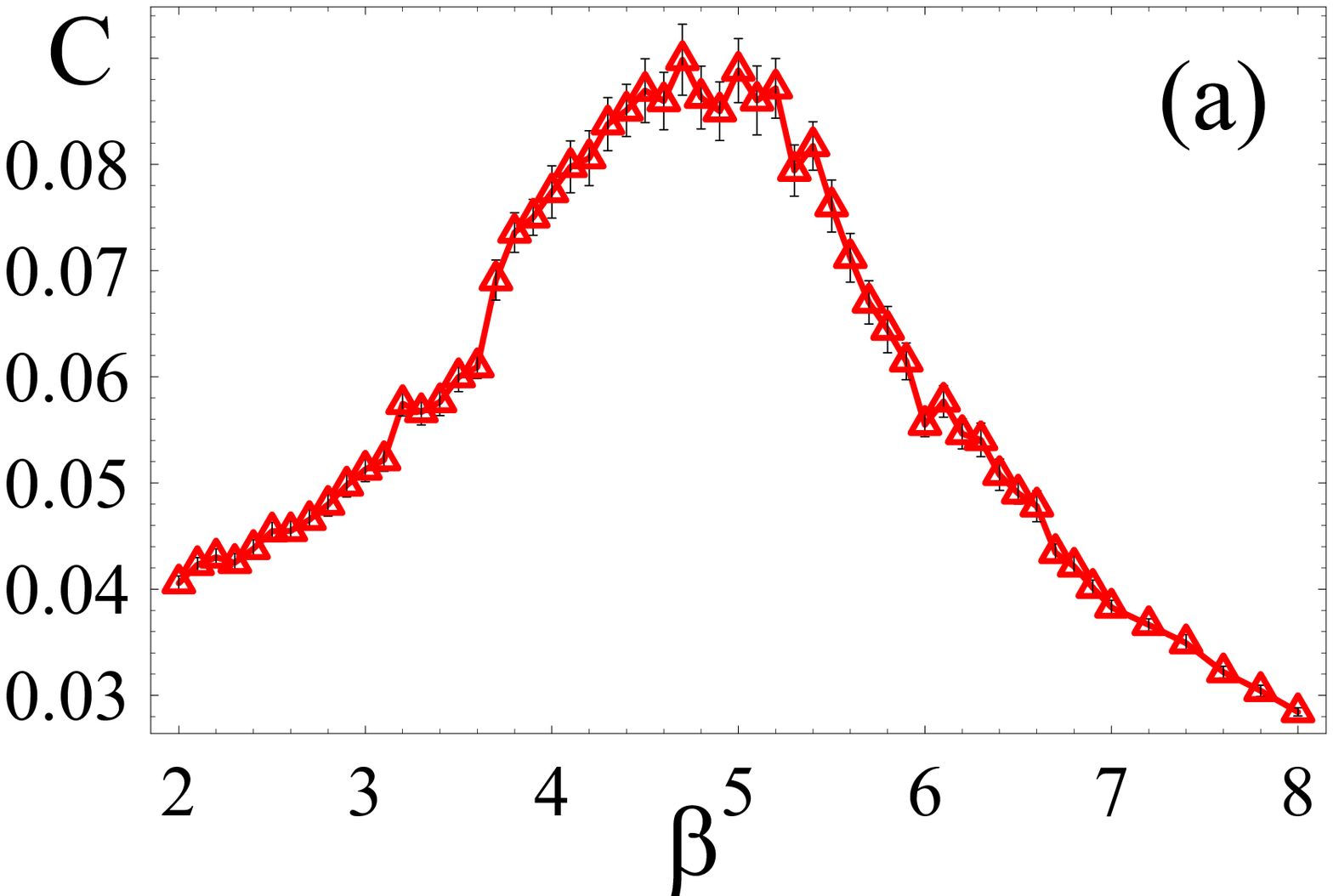}
\hspace{-0.0cm}
\includegraphics[width=4.1cm]{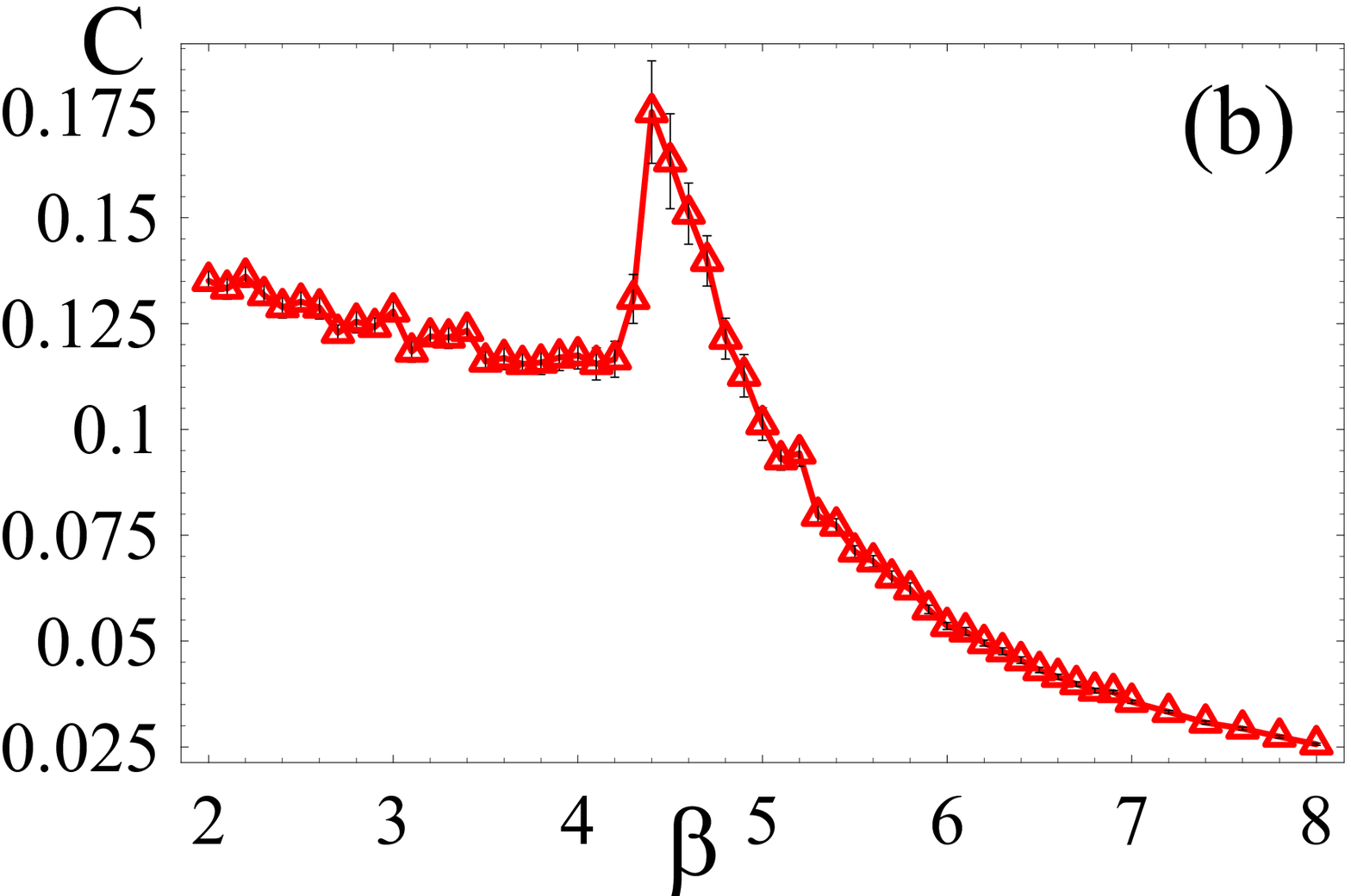}

\hspace{-0.2cm}
\includegraphics[width=4.1cm]{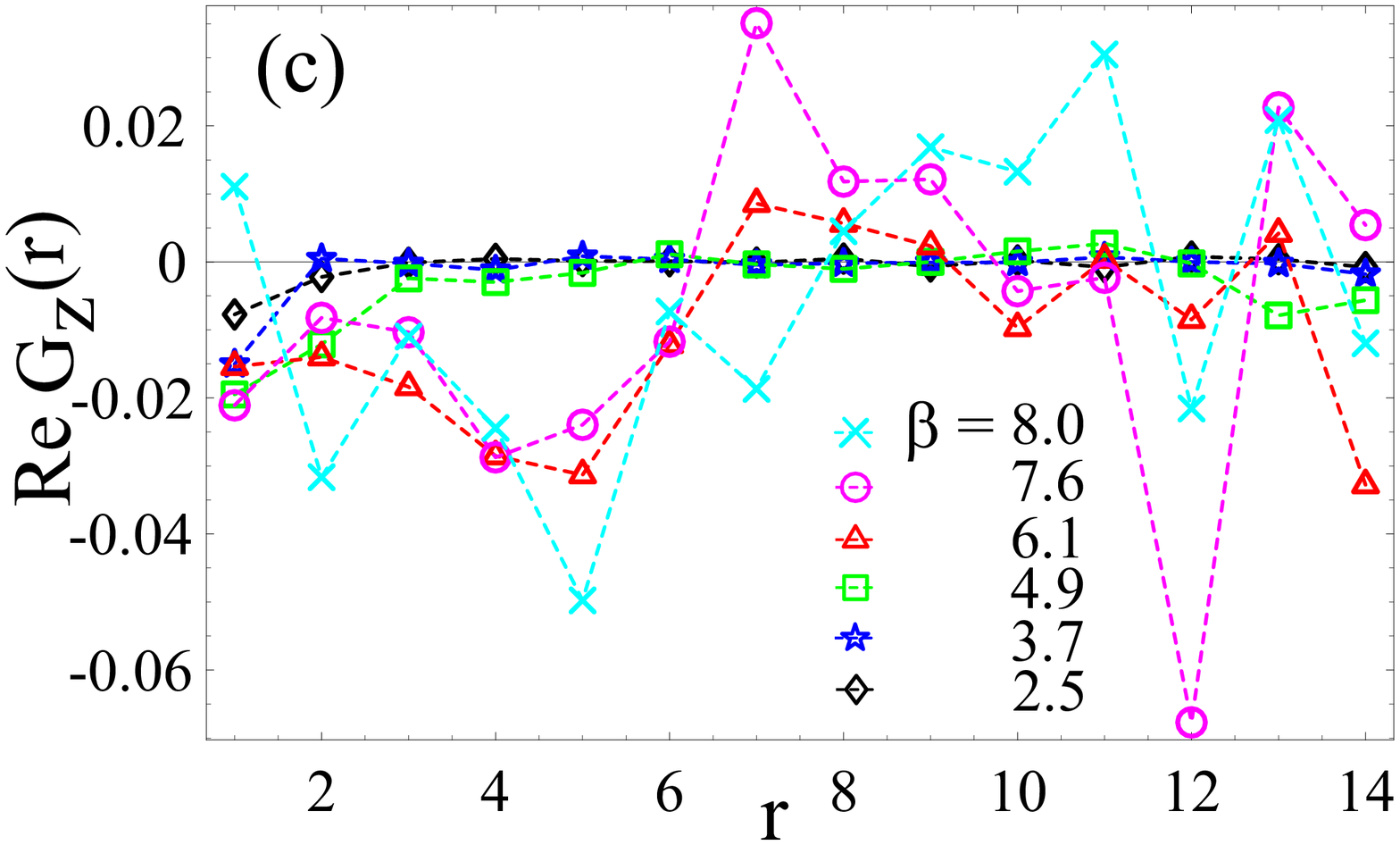}
\hspace{0.1cm}
\includegraphics[width=4.1cm]{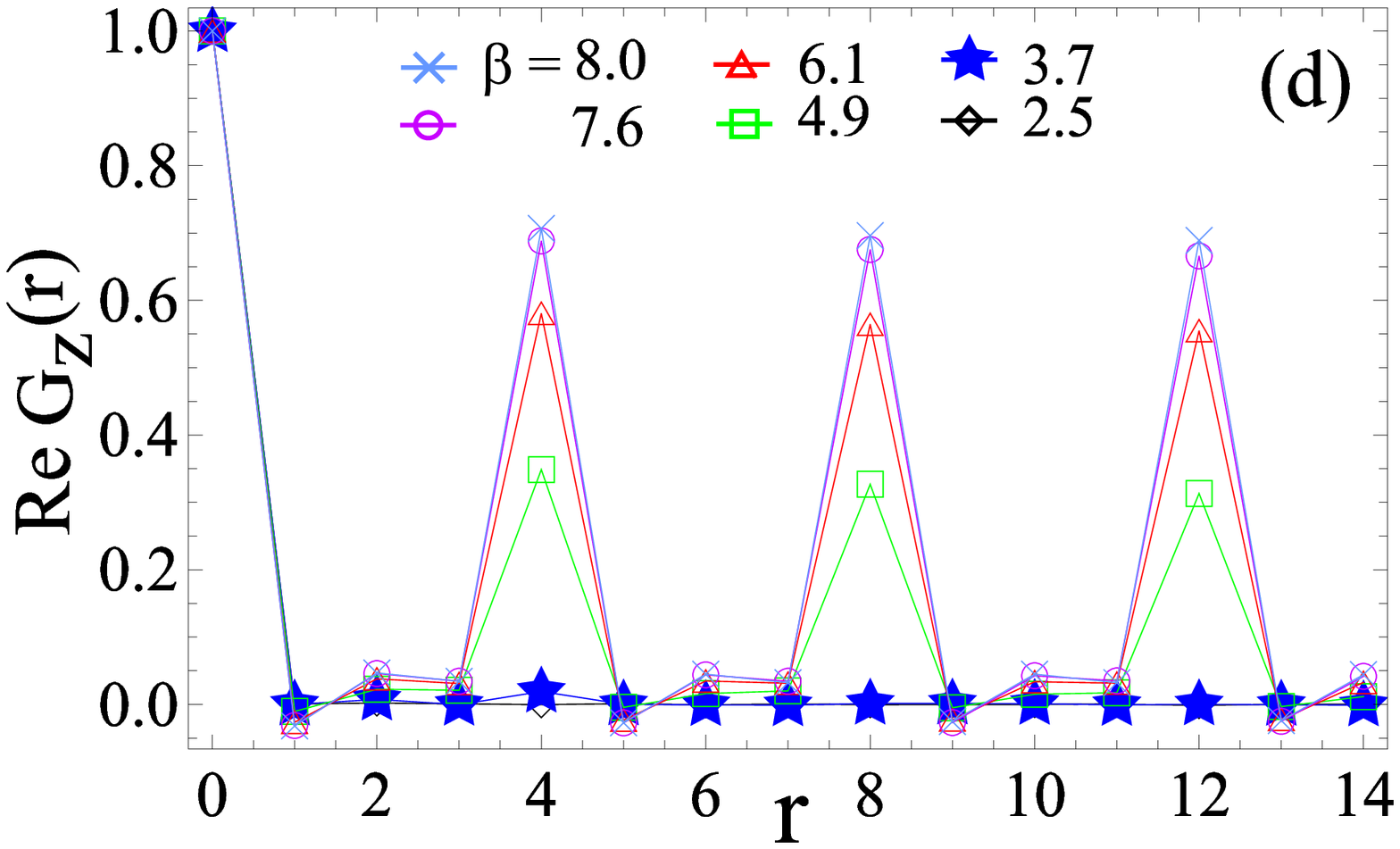}
\end{center}
\vspace{-0.5cm}
\caption{
The results for a constant magnetic field 
$\vec{m}_x=(0,0,m_{x3})^{\rm t}$ 
at $c_1=c_3=0.2$ 
($L=16$).
(a,b) Specific heat $C$ 
and (c,d) the real part of SC correlation function $G_z(r)$
in the 1-2 plane.
(a,c) $m_{x3}=\pi/4$ and (b,d) $m_{x3}=\pi$.
}
\label{cmg}
\end{figure}

In Fig.\ref{cmg}, we show $C$ and $G_z(x)$ for two cases of fixed $\vec{m}_x$.
For the case with 
$\vec{m}_{x}=(0,0,{\pi \over 4})^{\rm t}$,
$C$ has a shape  similar to $C_1$ in
Fig.\ref{C:magsc}c, and indicates a SC phase transition at 
$\beta \simeq 4.8$.
$G_z(r)$  
exhibits fluctuating behavior even for low $T$'s, $\beta \geq 4.9$.
This suggests that vortices are spontaneouly generated in the SC state
violating spatial unifomity, and their locations fluctuate.
In the other case of $\vec{m}_{x}=(0,0,\pi)^{\rm t}$, 
$C$ has a  sharper peak at $\beta\simeq 4.5$, and 
$G_z(r)$ exhibits clear periodically oscillating behavior with the period 
$4\times$(lattice spacing).
This implies that, in this case, locations of vortices are rather
stable compared with the case of $m_{x3}={\pi \over 4}$.

\begin{figure}[b]
\begin{center}
  \includegraphics[width=40mm]{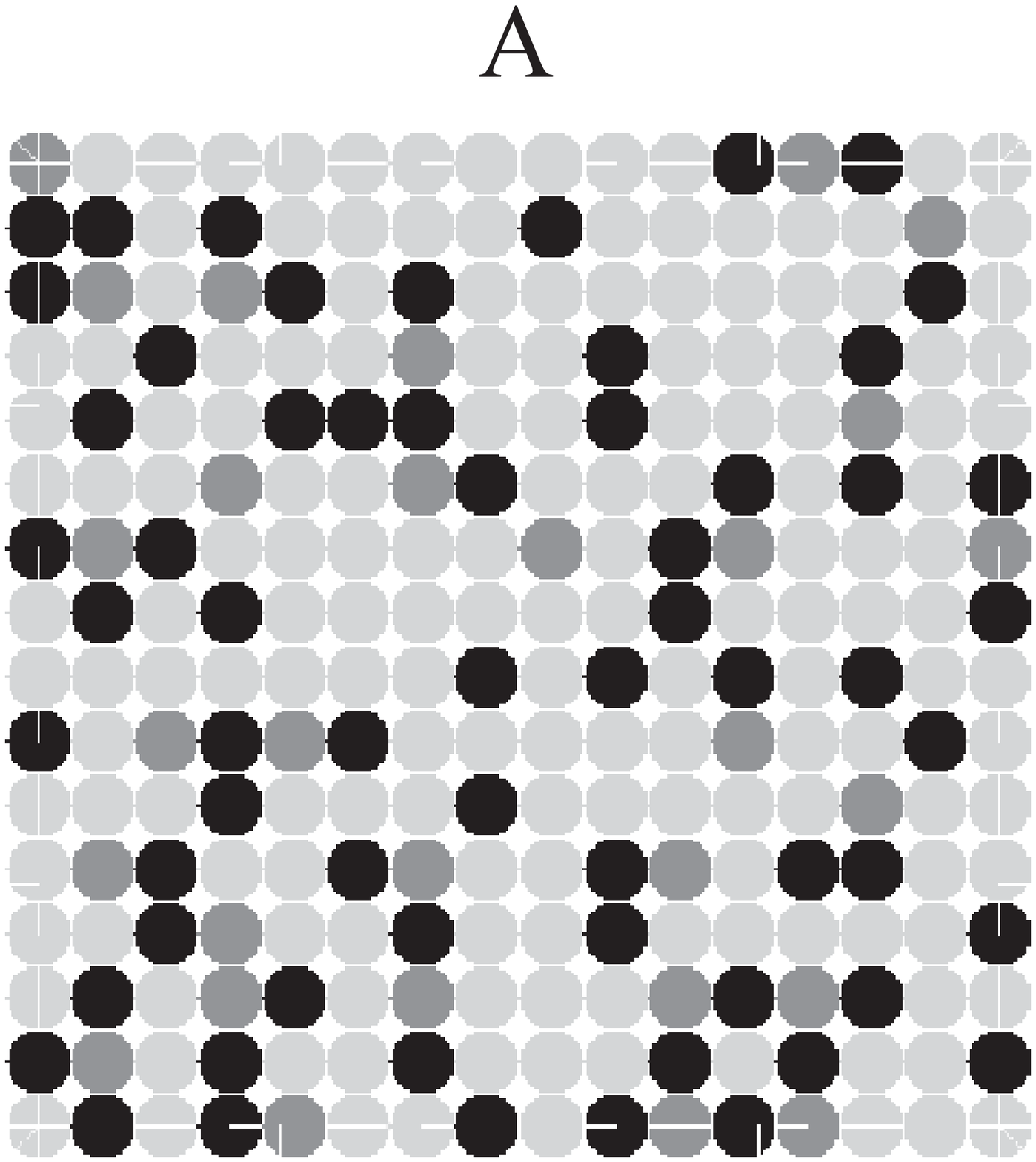}
  \includegraphics[width=40mm]{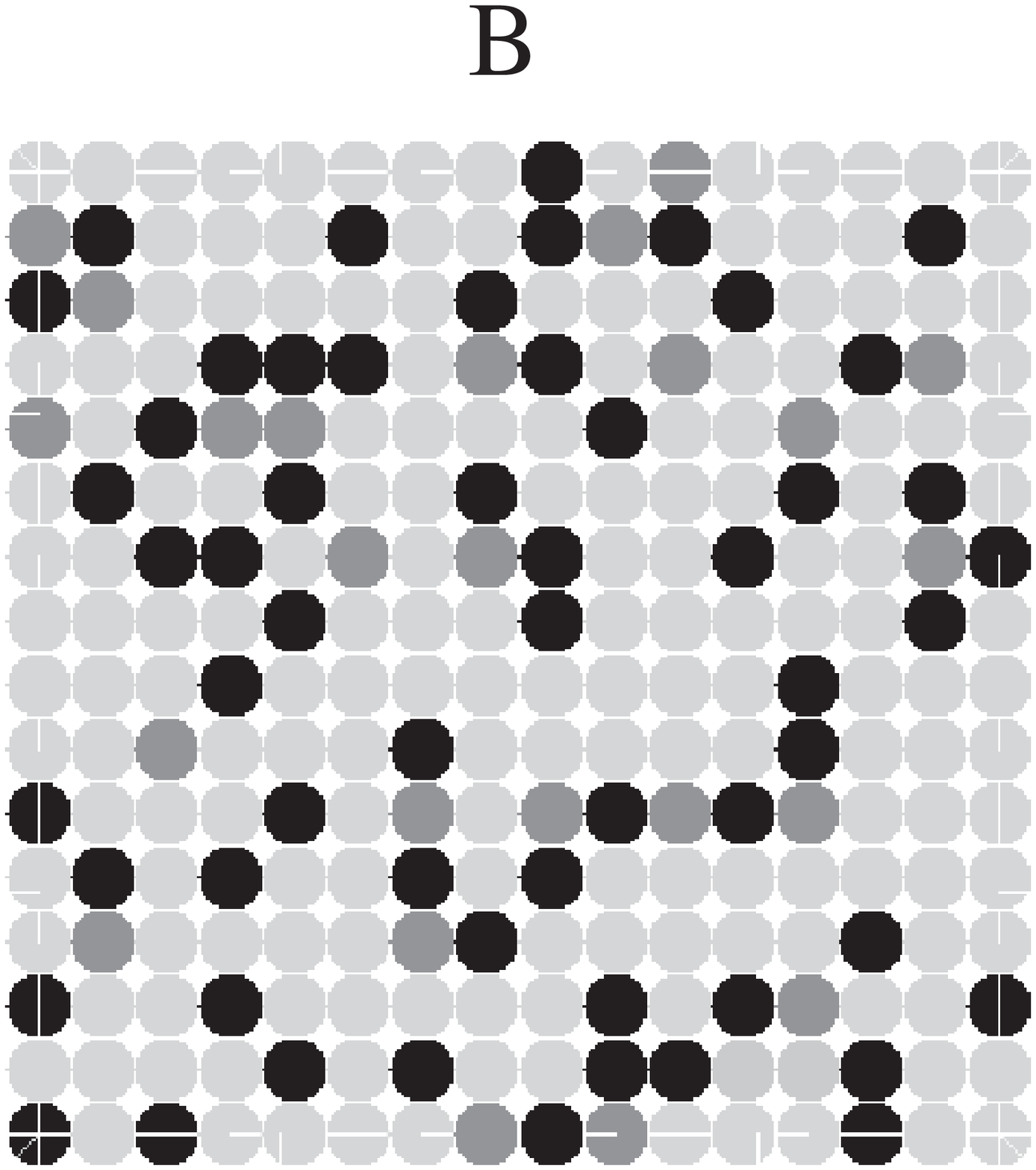}\\
\vspace{-0.7cm}
  \includegraphics[width=40mm]{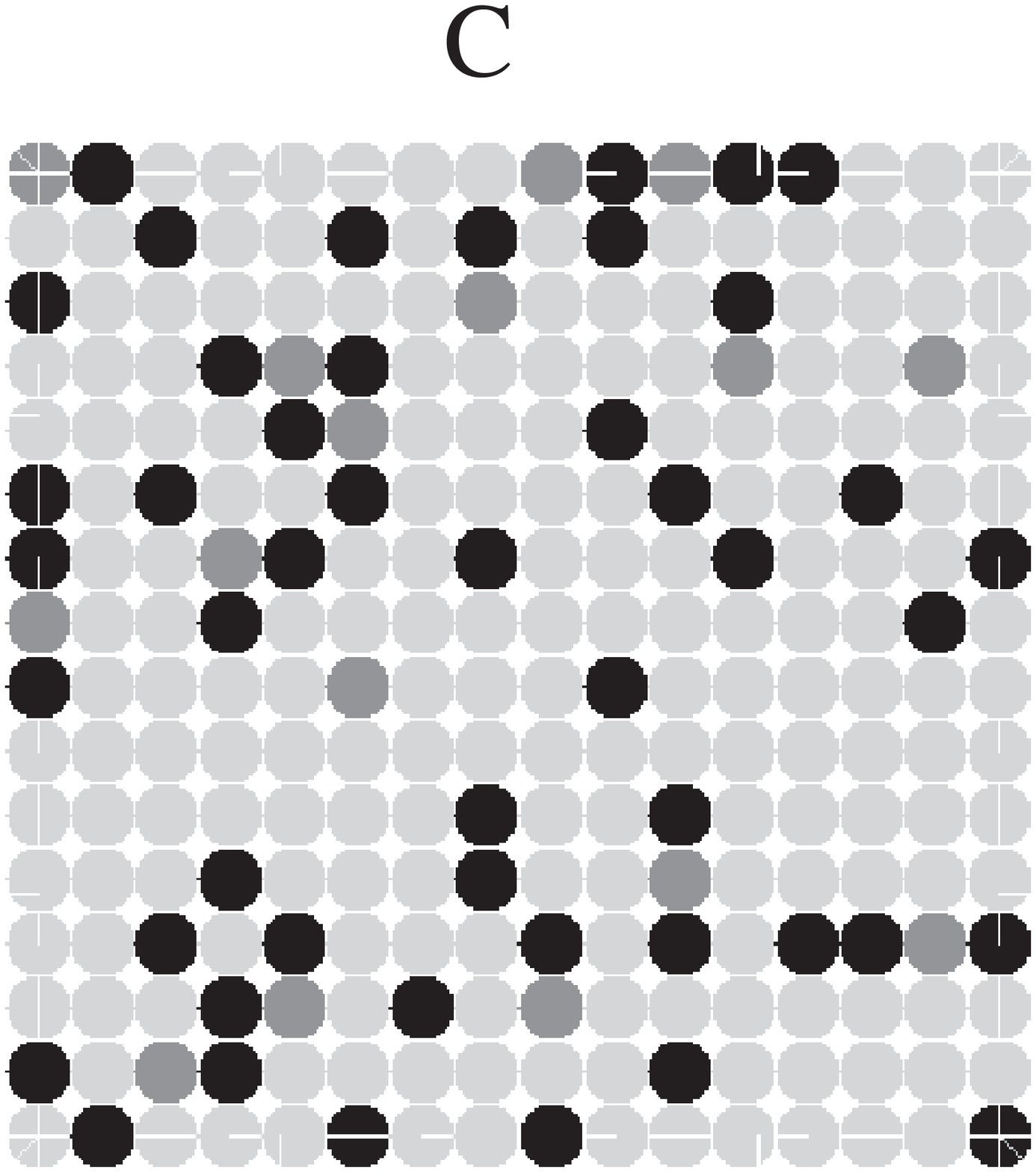}
  \includegraphics[width=40mm]{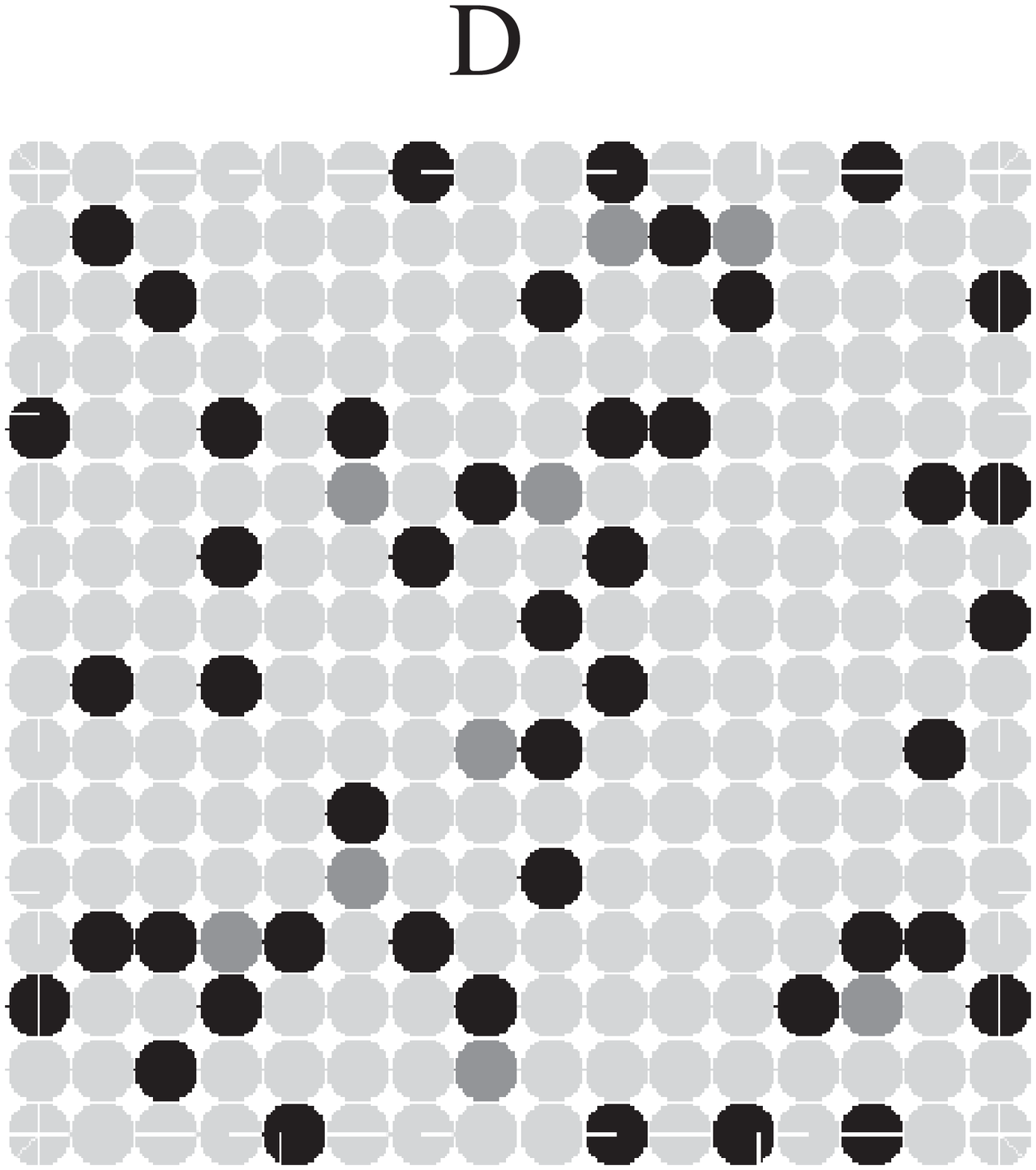}\\
\vspace{-0.3cm}
$V_x^\pm$ \\
  
\includegraphics[width=17mm]{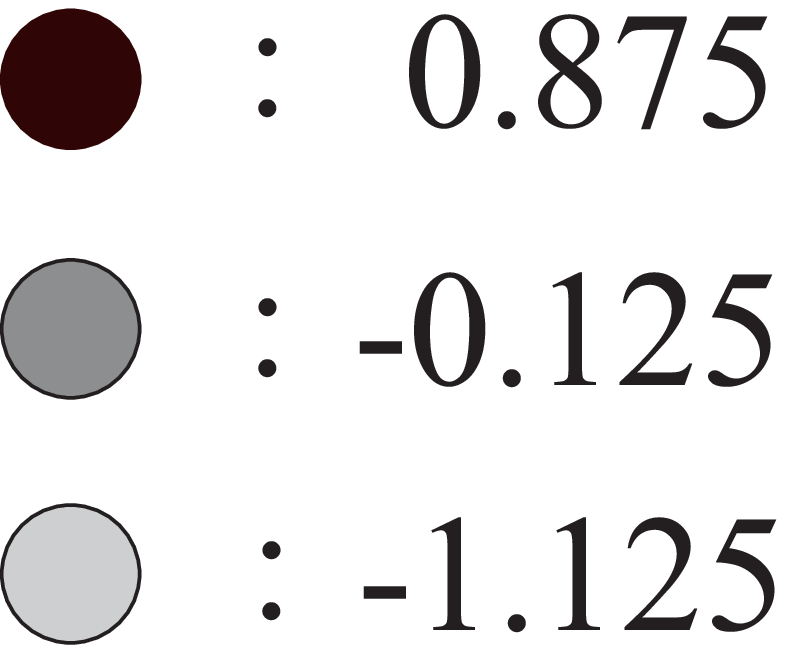}
\end{center}
\vspace{-0.3cm}
\caption{
Snapshots of vortex densities $V_x^{\pm}$ at $c_1=c_3=0.2$ 
for a fixed $\vec{m}=(0,0,\pi/4)\ (L=16)$.
Black dots; $V_x^{\pm}$= 0.875, Dark gray dots; -1.125, Light gray
dots; -0.125. 
(A)$V_x^+$ at $\beta=3.0$, (B)$V_x^-$ at $\beta=3.0$,
(C)$V_x^+$ at $\beta=7.0$, (D)$V_x^-$ at $\beta=7.0$.
The average magnitude $\la |V_{x\pm} | \ra$ is 
(A) 0.387, (B) 0.380, (C) 0.331 and (D) 0.335.
The points  $V_x^{\pm}=-0.125=-m_3/(2\pi)$ reflect $\vec{m}$ itself.}
\label{snapvor1}
\end{figure}

In order to verify the above expectation, we calculate vortex density
directly.
In the present model, one may define the  following two kinds of 
gauge-invariant vortex
densities $V_x^+$ and $V_x^-$ in the 1-2 plane;
\be
z_x^{\pm}&\equiv& z_{x1}\pm iz_{x2}\equiv \rho_x^{\pm}
\exp(i\theta_x^{\pm}),\nn
 V_x^{\pm} &\equiv& \frac{1}{2\pi}
[{\rm mod}(\theta_{x+1}^{\pm}-\theta_x^{\pm}-A_{x1}) \nn
&&+{\rm mod}(\theta_{x+1+2}^{\pm}-\theta_{x+1}^{\pm}-A_{x+1,2}) \nn
&&-{\rm mod}(\theta_{x+1+2}^{\pm}-\theta_{x+2}^{\pm}-A_{x+2,1}) \nn
&&-{\rm mod}(\theta_{x+2}^{\pm}-\theta_x^{\pm}-A_{x2})],
\ee
where mod$(x)\equiv$ mod$(x,2\pi)$. In short, $V_{x}^{\pm}$ 
describes vortices of electron pairs with the amplitude
$\psi_{\uparrow\uparrow({\downarrow\downarrow})}
=\psi_1\pm i\psi_2\propto z_1\pm iz_2$.

\begin{figure}[t]
\begin{center}
   \includegraphics[width=25mm]{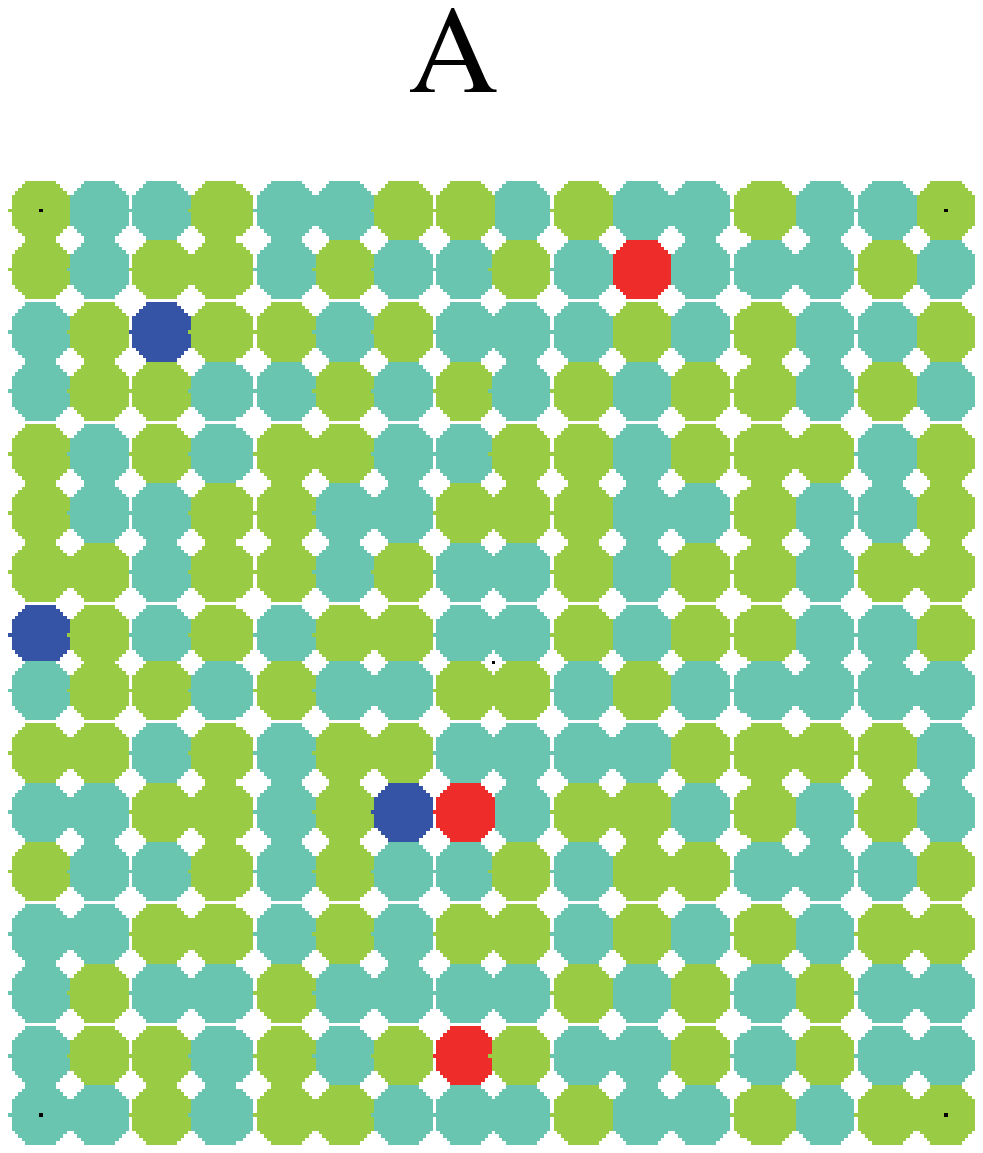}
\hspace{1cm}
  \includegraphics[width=25mm]{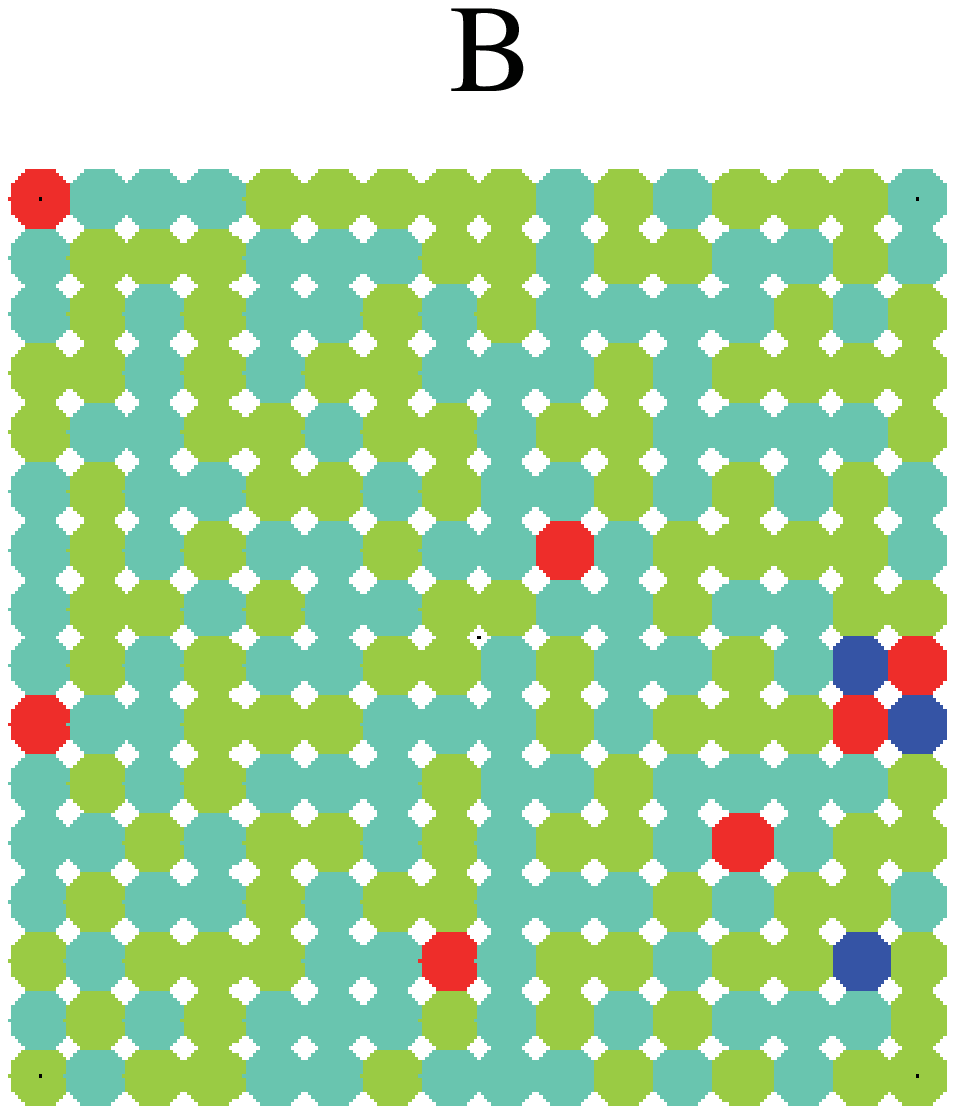}\\
\vspace{0cm}
  \includegraphics[width=25mm]{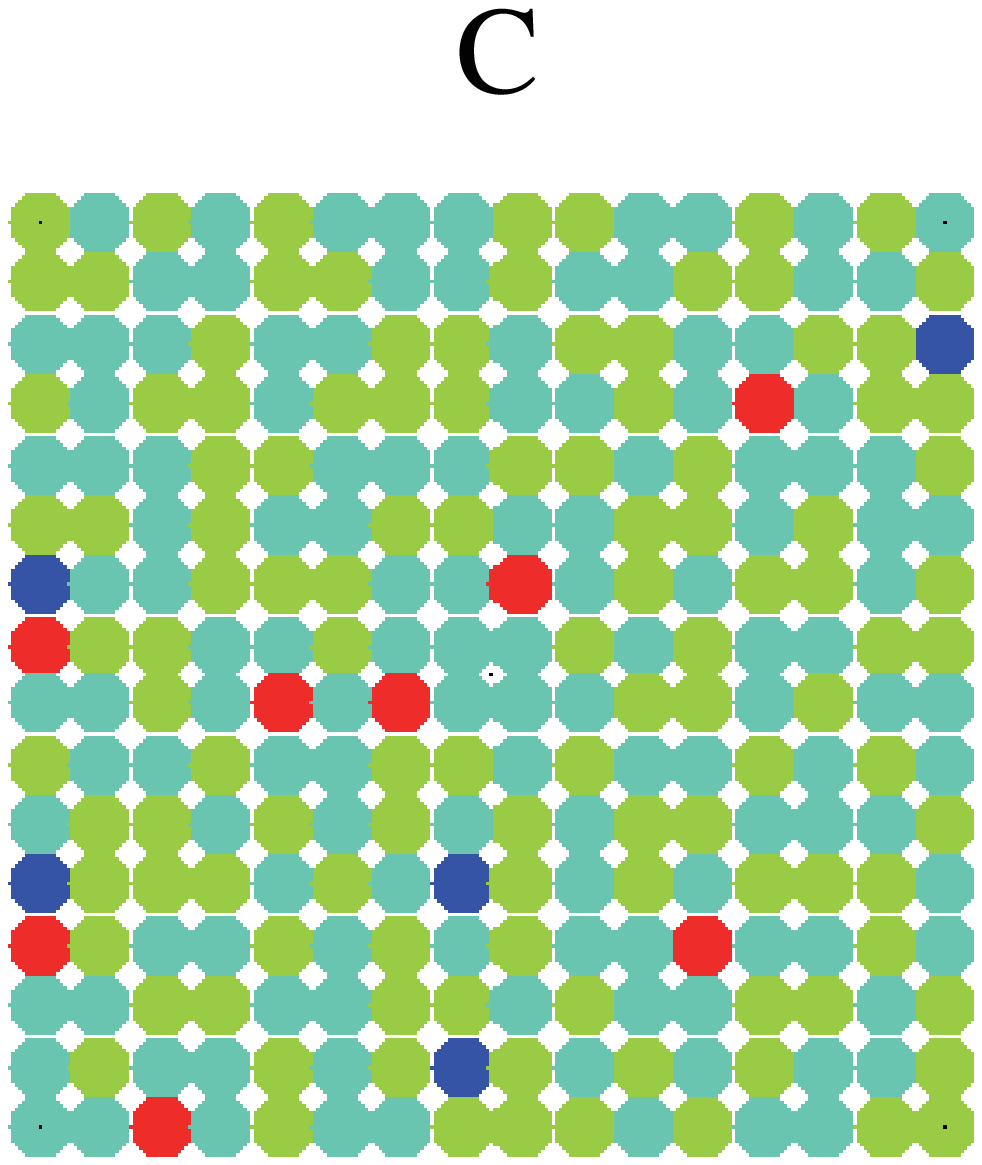}
\hspace{1cm}
  \includegraphics[width=25mm]{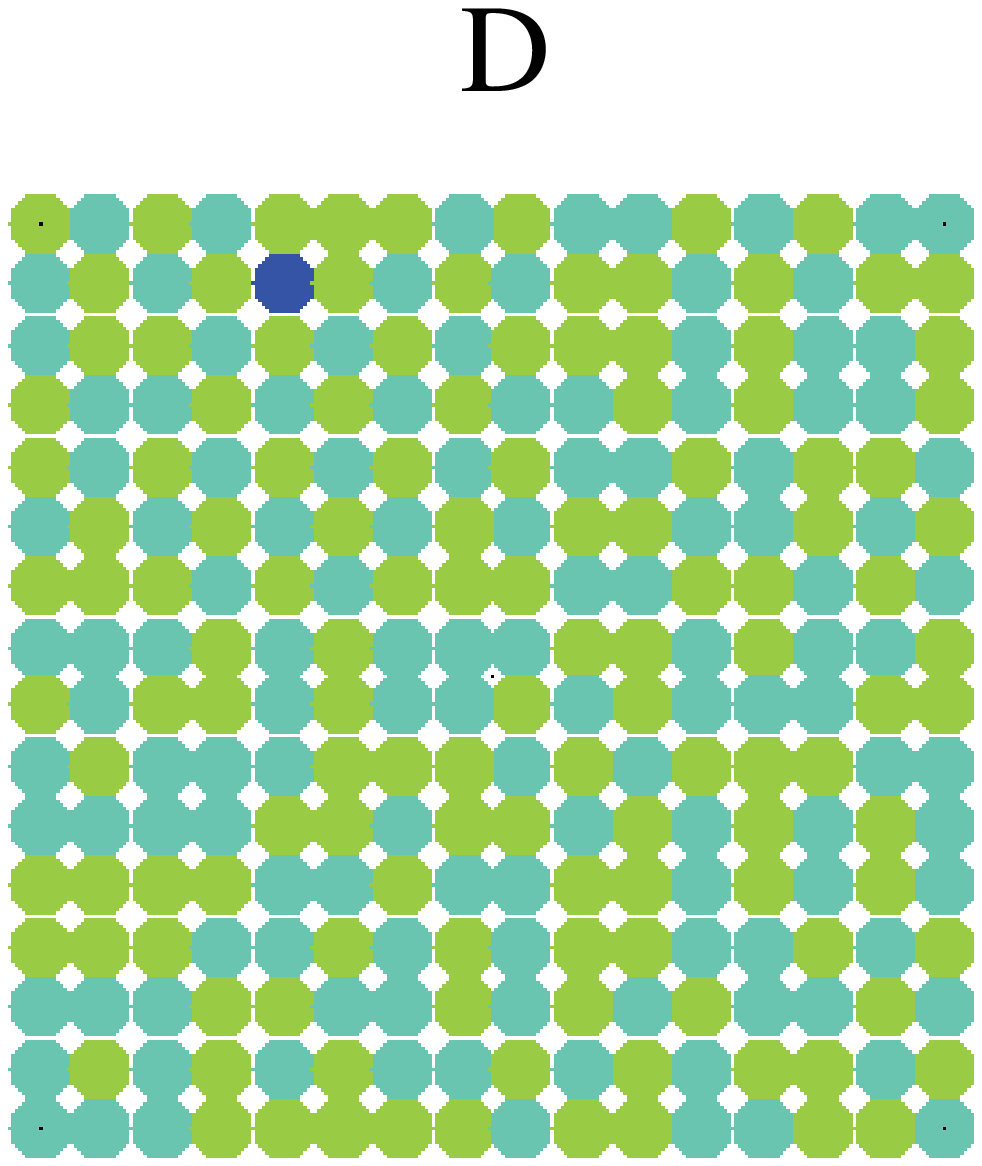}\\
 $V_x^{\pm}$ \\
  \includegraphics[width=12mm]{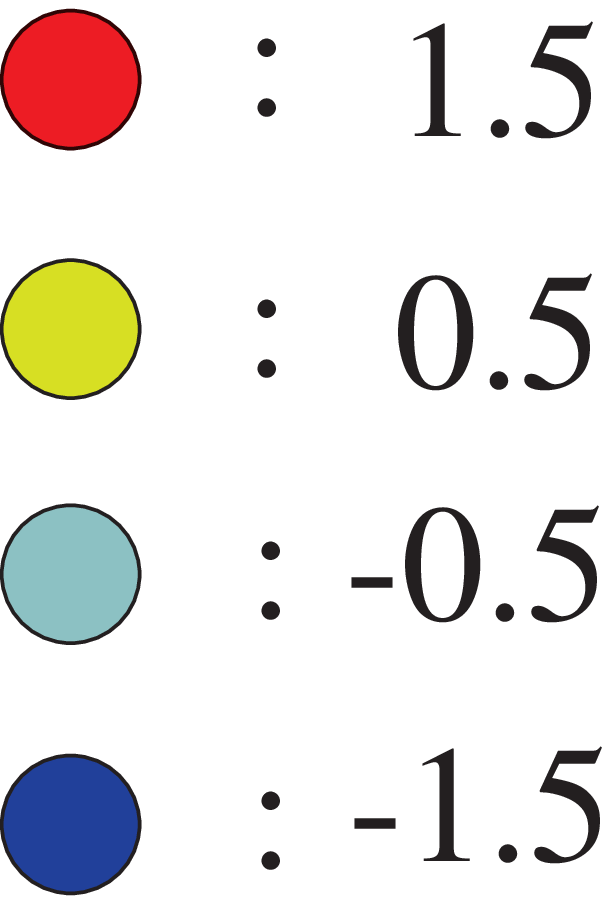}
\end{center}
\vspace{-1.5cm}
\caption{
Snapshots of vortex densities $V_x^{\pm}$ at $c_1=c_3=0.2$ 
for a fixed $\vec{m}=(0,0,\pi)\ (L=16)$.
(A)$V_x^+$ at $\beta=3.0$, (B)$V_x^-$ at $\beta=3.0$,
(C)$V_x^+$ at $\beta=7.0$, (D)$V_x^-$ at $\beta=7.0$.
The average magnitude $\la |V_x^{\pm} | \ra$ is 
(A) 0.528, (B) 0.537, (C) 0.560 and (D) 0.508.
The points  $V_x^{\pm}=-0.5=-m_3/(2\pi)$ reflect $\vec{m}$ itself.}
\label{snapvor2}
\end{figure}

In Fig.\ref{snapvor1} we present snapshots of $V_x^{\pm}$ at 
$c_1=c_3=0.2$ 
for fixed values of
gauge potential $A_{x\mu}$ corresponding to a constant magnetization,
$\vec{m}=(0,0,\frac{\pi}{4})^{\rm t}$. 
It shows that
(i) both of the fluctuations around zero, $\la |V_x^{\pm} |\ra$,
decrease as $\beta$ increases, and (ii)
$V_x^+$ has larger fluctuations than $V_x^-$ at high $T$,
whereas smaller ones at low $T$.
These behaviors are consistent with the Zeeman $c_3$-term in the
energy $f_x$ of (\ref{FLGL}), which distinguishes the $z_x^+$ order and 
the $z_x^-$ order, and the fact that $\vec{m}$ directs to 
the third-direction in the present case. 
In Fig.\ref{snapvor2}, we also show the vortex snapshots at $c_1=c_3=0.2$ 
and $\vec{m}=(0,0,\pi)^{\rm t}$.
Compared with the case $\vec{m}=(0,0,\frac{\pi}{4})^{\rm t}$,
vortices here are located rather systematically as we expected from the
result of correlation function $G_z(r)$.

\begin{figure}[b]
\begin{center}
\includegraphics[width=4.2cm]{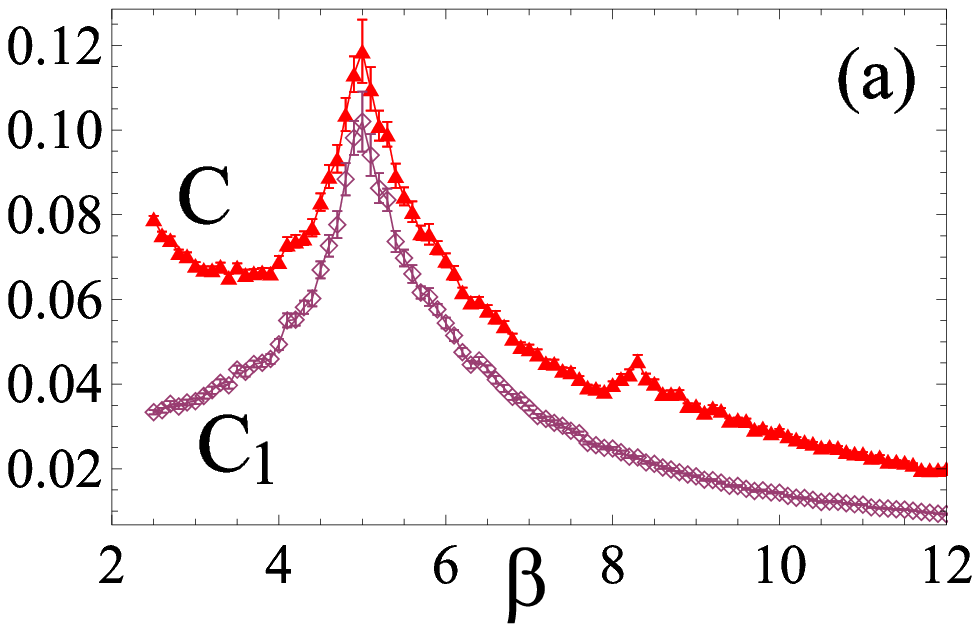}
\includegraphics[width=4.2cm]{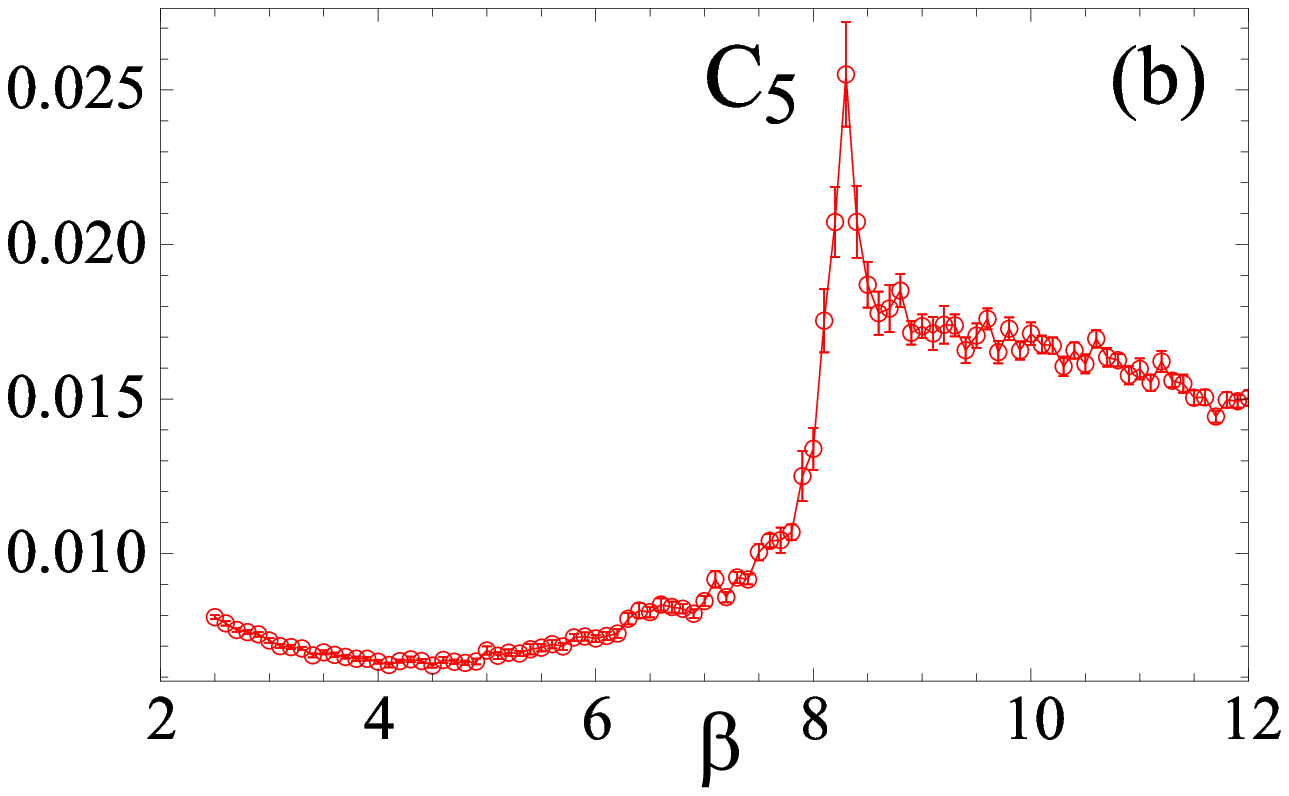}
\end{center}
\vspace{-0.5cm}
\caption{
Specific heats for $c_5=0.4\ (L=12)$.
(a) Total $C$ and the specific heat $C_1$ of the $c_1$-term,
(b) $C_5$ of the $c_5$-term.
Order of two phase transitions is interchanged ($\beta_{\rm SC} < \beta_{\rm FM}$). 
}
\label{c5low}
\end{figure}

\subsection{Order of FM and SC phase transitions and phase diagram}

Let us next examine the possibility that the order of 
the FM and SC phase transitions are interchanged
as the value of $c_5$ is decreased.
As the $c_5$-term in $f_x$ tends to align $\vec{m}_x$,
$T_{\rm FM}$ decreases as $c_5$ is decreased.
Most of the FMSC materials loses the FM order as the applied $P$ 
is increased, and then it is phenomenologically expected that $c_5$
is a decreasing function of $P$. 
We consider several cases with $c_5=1.5, 1.0, 0.7, 0.5, 0.4$ and $0.3$,
while other $c_i$ are fixed to the same values as those in Fig.\ref{C:magsc} where
$c_5=1.0$. 

We find that  the order of two phase transitions actually interchanges 
at $c_5\simeq 0.5$.
In Fig.\ref{c5low}, we show the  
specific heat $C, C_1, C_5$ for $c_5=0.4$. $C$ has the two peaks at 
$\beta_{\rm SC}\sim 5.0 < \beta_{\rm FM} \sim 8.3$.
$C_1$ is sharper than in the case of $c_5=1.0$ in Fig.\ref{C:magsc}.

\begin{figure}[t]
\vspace{-2.8cm}
  \hspace{-1.9cm}
 \begin{minipage}{0.45\hsize}
  \begin{center}
\includegraphics[width=6.6cm]{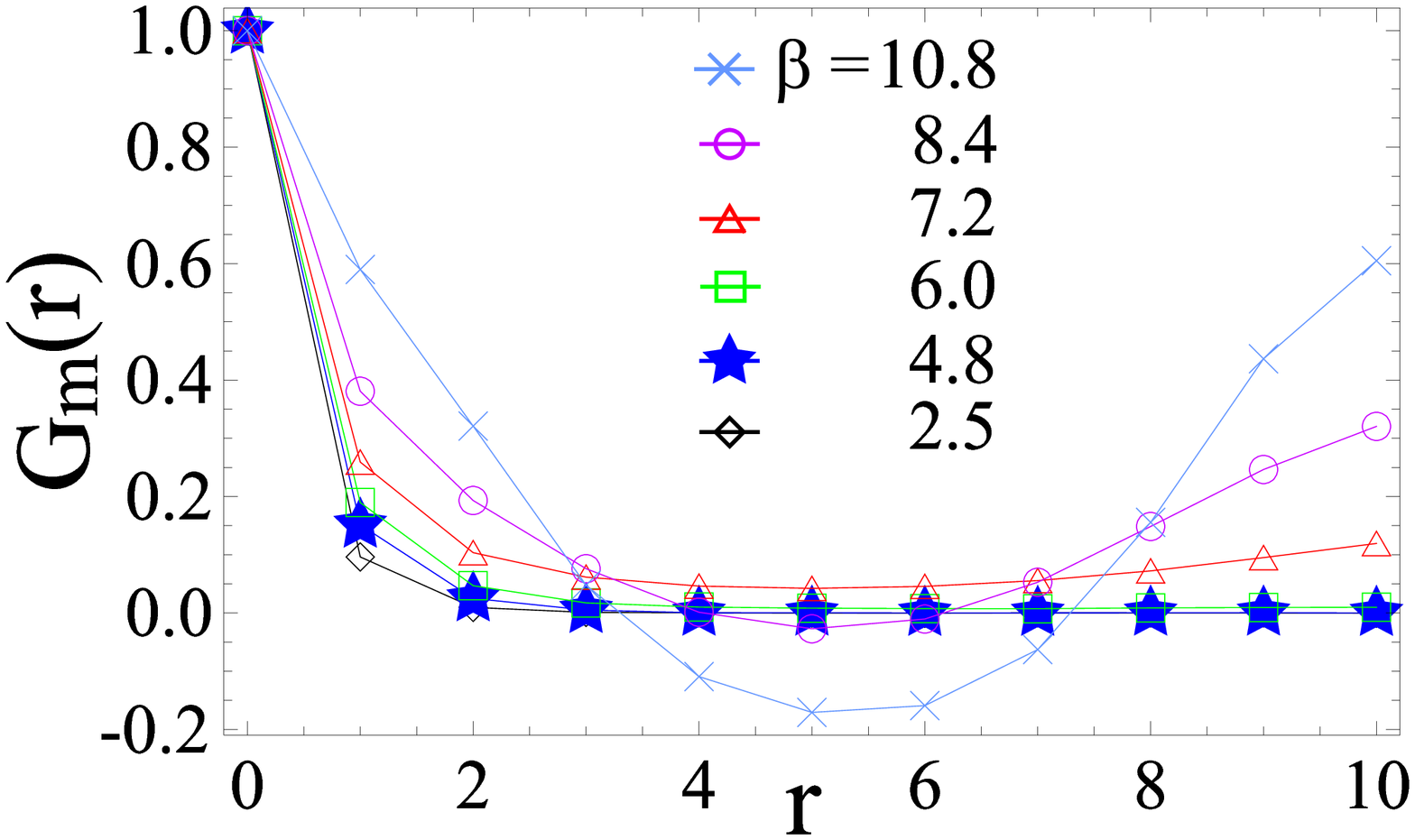}
  \end{center}
 \end{minipage}
     \hspace{1.8cm}
 \begin{minipage}{0.45\hsize}
    \vspace{2.1cm}
  \begin{center}
\includegraphics[width=4.3cm]{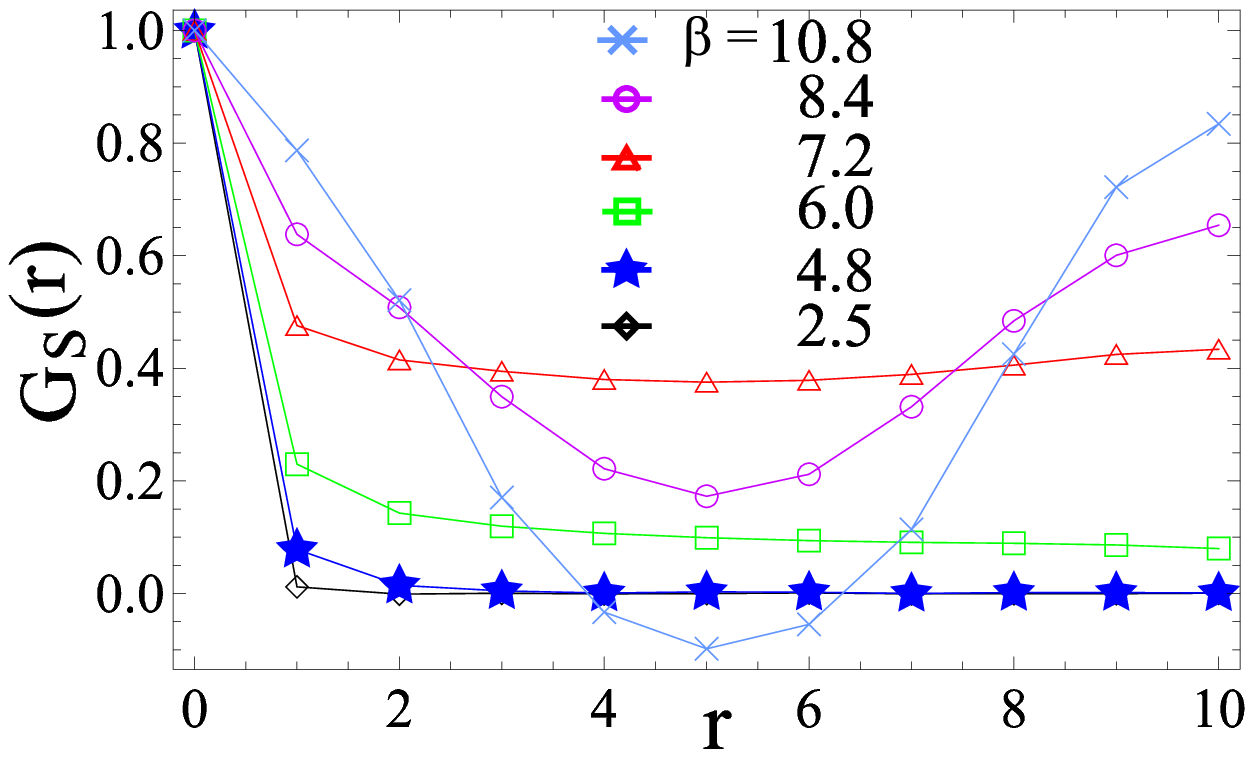}
  \end{center}
 \end{minipage}
 \vspace{-0.6cm}
\caption{
Correlation functions $G_{m}(r)$ and 
 $G_{S}(r)$ at $c_5=0.4$ ($L=12$). The other $c_i$
are the same as those in Fig.\ref{C:magsc}. 
They exhibit orders near the surface of the lattice. 
}
\label{c5lowcor}

\end{figure}

In Fig.\ref{c5lowcor} we present $G_{m}(r)$ and 
$G_{S}(r)$ with $r$ in the 1-2 plane for $c_5=0.4$, 
which  exhibit very peculiar behavior;
In {\em the FM and SC coexisting phase} ($T < T_{\rm FM}$), 
they have nonvanishing
values only near the surface of the lattice  
in contrast with Fig.\ref{correlation1}.
This behavior survives in larger systems.
For example, we define the thickness $\Delta L$ of the coexisting region 
such that the ordered region in the 1-2 plane occupies the interval  
$2+\Delta L +({\rm disordered\ region}) + \Delta L+2$
in the lattice length $2+L+2$ in the $\mu=1,2$ directions.
We obtain  $\Delta L \simeq 3$ for $L=12$(Fig.\ref{c5lowcor}) 
and $\Delta L \simeq 4$ for $L=16$, so about the outer half region in 
the linear dimension is occupied by the ordered state.
This implies that the FMSC coexisting phase appear
in the region including the surface of the material, 
and not in the central region inside the system. 
We note that this ``surface" region is not two-dimensional but 3D,
because the SC-FM transition is a genuine second-order one,
which is not allowed in a two-dimensional system\cite{mermin}.
This phenomenon is a prediction of the present model.

\begin{figure}[b]
\begin{center}
\hspace{-0.2cm}
\includegraphics[width=4cm]{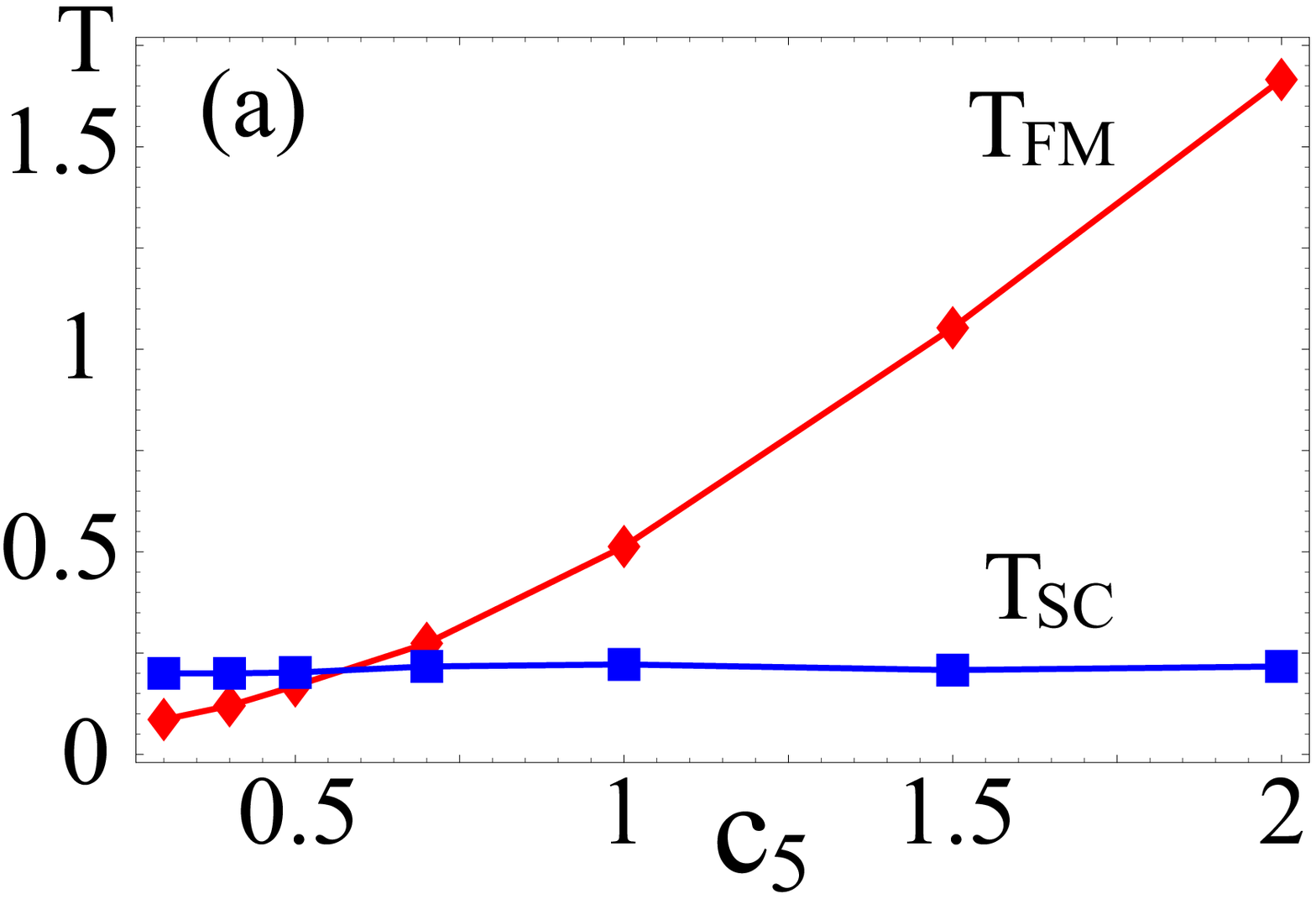}
\hspace{0.3cm}
\includegraphics[width=4.2cm]{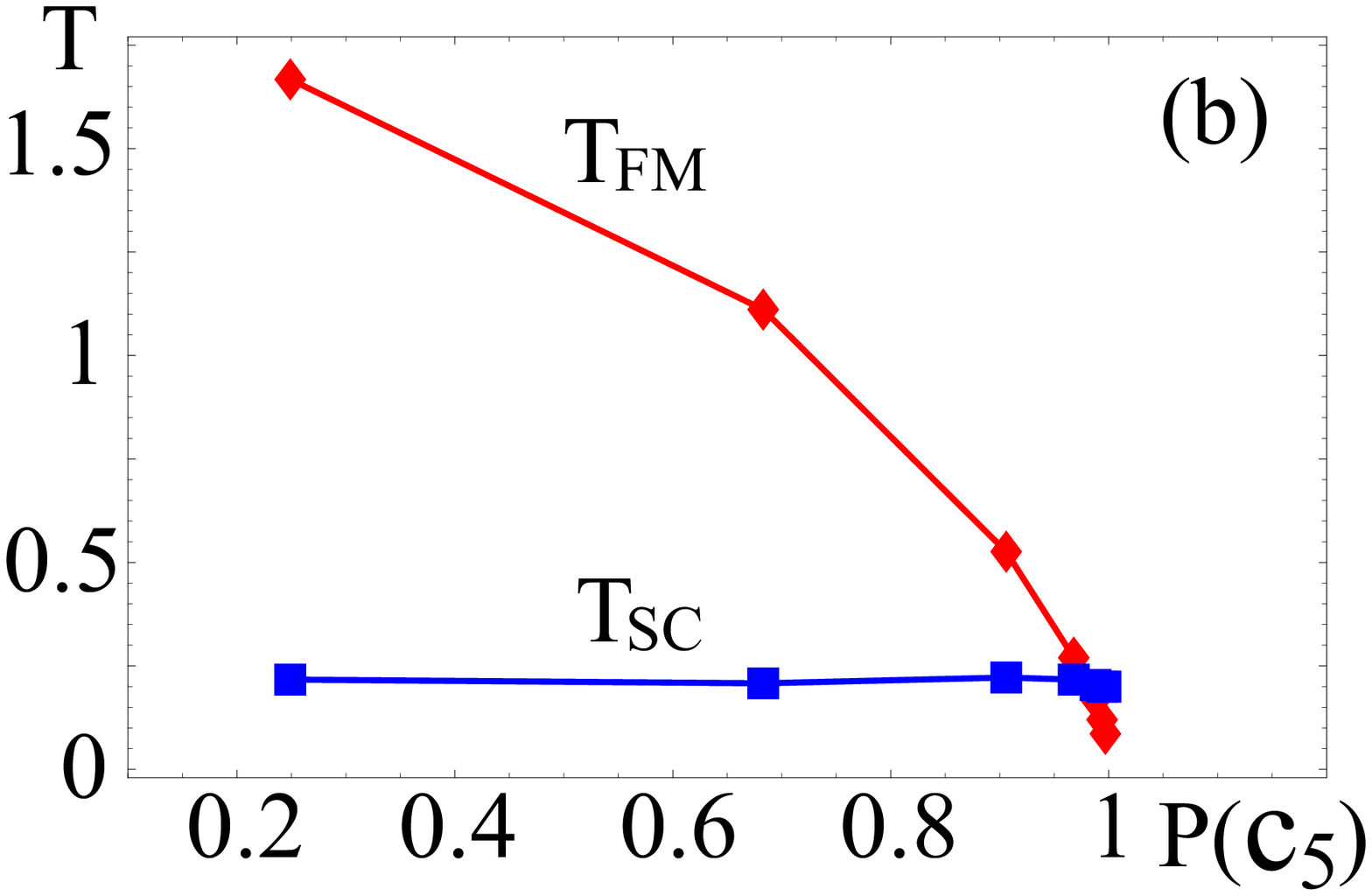}
\end{center}
\vspace{-0.5cm}~\caption{
Phase diagrams in (a) $c_5$-$T$ and (b) $P$-$T$ planes with
$P_{\rm c}=1.0$, $c_5^\star=2.2$ and $\gamma=3.0$.
$c_1-c_4$ are same as in Fig.\ref{C:magsc}.\\
}
\label{PDP}
\end{figure}

It is intriguing to draw a phase diagram in the $P$-$T$ plane
assuming certain phenomenological relation between $c_5$ and $P$.
In the experiments,
the critical temperature $T_{\rm FM}$ is a decreasing function of $P$.
This means that the parameters $c_2,\ c_4$ and $c_5$ in Eq.(\ref{FLGL})
vary as functions of $P$.
Changes of $c_2$ and/or $c_4$ influence the magnitude of the 
magnetization vector $\vec{m}_x$ and result in a change of $c_5$
after a renormalization of $\vec{m}_x$.
Then for example, one may ``phenomenologically" assume
\be
c_5=c^\star_5\Big(1-{P(c_5) \over P_{\rm c}}\Big)^{1/\gamma},
\label{c5P}
\ee
where $P_{\rm c}$ is the critical pressure at which the FM order
disappears even at $T \rightarrow 0$ (i.e., at $c_5=0$), 
$c_5^\star$ is the value of $c_5$ at which $P=0$,
and the power $\gamma$ is a fitting parameter.
In Fig.\ref{PDP} we show the phase diagram drawn with certain choice
of these parameters.
This phase diagram has a similar structure to the experimental
result of UCoGe.

In the same way as decreasing $c_5$, the case of increasing 
$c_3$  has been studied with several choices of $c_5$ and  
$(c_1,c_2,c_4)=(0.2,0.5,4.0)$.
Both $T_{\rm FM}$ and $T_{\rm SC}$ increase as $c_3$ increases.
Furthermore, for sufficiently large values 
such as $c_3=1.5$, 
$T_{\rm FM} > T_{\rm SC}$ even for $c_5=0.5$ as expected.
This indicates  
that the present model with larger $c_3$ 
may provide a phase diagram similar to that of UGe$_2$ and URhGe.

\section{Conclusion}

In summary, we have proposed a GL model defined on the 3D lattice for
the FMSC state, and shown that it explains some experimental observations
such as the

\nin
phase diagram and 
the homogeneous and inhomogeneous FMSC states.
This model naturally includes effects of 
topological excitations, vortices, 
that play an important role for the SC phase transition in the FM state.
Although the obtained global phase structure is similar to that
of MFT, the appearance of inhomogeneous configurations
such as the FMSC state and vortex configurations are certainly beyond the
scope of MFT.   
In the present analysis, we consider the ``London limit", in which the 
radial fluctuations of the two-gap SC order parameters are ignored.
As we explained in Sec.II.A., these fluctuations may change the order of SC
phase transitions and may play an important role 
in SC transitions that are induced by an external 
magnetic-field.
This problem is under study and results will be reported in a 
future publication.

\vspace{0.8cm}
\acknowledgments
We thank Kenji Sawamura, Kento Uchida, and Tomonori Shimizu
for their collaborations in the early stage of the present study.
This work was partially supported by Grant-in-Aid
for Scientific Research from Japan Society for the 
Promotion of Science under Grant No.20540264 and No23540301.





\begin{thebibliography}{99}
\bibitem{2e}
We have included the charge $-2e$ in $A_{x\mu}$.
\end{thebibliography}
\end{document}